\documentclass[graybox]{svmult}

\usepackage{mathptmx}       
\usepackage{helvet}         
\usepackage{courier}        
\usepackage{type1cm}        
\usepackage{makeidx}         
\usepackage{graphicx}        
\usepackage{multicol}        
\usepackage[bottom]{footmisc}

\usepackage{amsmath,amsfonts}
\usepackage{array}
\usepackage{graphicx}
\usepackage{multicol}
\usepackage[english]{babel}
\usepackage[figurename={},tablename={}]{caption}
\usepackage{float}

\usepackage{mathrsfs}
\usepackage[Symbol]{upgreek}
\usepackage{bbm}
\usepackage{dsfont}
\usepackage{amssymb}
\usepackage{wasysym}
\usepackage[pdf]{pstricks}
\usepackage{colortab}
\usepackage{url} 
\usepackage[off]{auto-pst-pdf}

\usepackage{feynmf}
\newcommand{\be}{\begin{equation}} 
\newcommand{\ee}{\end{equation}} 
\newcommand{\bea}{\begin{eqnarray}} 
\newcommand{\eea}{\end{eqnarray}} 

\def\bs#1\es{\begin{split}#1\end{split}}
\def\ba#1\ea{\begin{align}#1\end{align}}
\def\bg#1\eg{\begin{gathered}#1\end{gathered}}

\def\half{{\frac12}}

\def\nn{\nonumber} \def\bd{\begin{document}} \def\ed{\end{document}}
\def\ds{\documentstyle} \let\fr=\frac \let\bl=\bigl \let\br=\bigr
\let\Br=\Bigr \let\Bl=\Bigl
\let\bm=\bibitem
\let\na=\nabla
\let\pa=\partial \let\ov=\overline

\def\del{\partial}
\def\vp{\varphi}
\def\st#1{{\scriptstyle #1}}
\def\sst#1{{\scriptscriptstyle #1}}
\def\sz{\scriptsize}
\def\oneone{\rlap 1\mkern4mu{\rm l}}
\def\td{\tilde}
\def\wtd{\widetilde}

\newcommand{\ra}{\rightarrow}
\newcommand{\lra}{\longrightarrow}
\newcommand{\Lra}{\Leftrightarrow}
\newcommand{\ap}{\alpha^\prime}
\newcommand{\bp}{\tilde \beta^\prime}
\newcommand{\tr}{{\rm tr} }
\newcommand{\Tr}{{\rm Tr} }

\renewcommand{\atop}[2]{\genfrac{}{}{0pt}{}{#1}{#2}}

\newenvironment{subitemize}{\begin{itemize}\scriptsize}{\end{itemize}}

\def\ie{{\it i.e.\ }}
\def\eg{{\it e.g.\ }}
\def\viz{{\it viz.}\ }
\def\nn{\nonumber}

\def\ft#1#2{\tfrac{#1}{#2}}
\def\fft#1#2{\frac{#1}{#2}}

\def\0{{\sst{(0)}}}
\def\1{{\sst{(1)}}}
\def\2{{\sst{(2)}}}
\def\3{{\sst{(3)}}}
\def\4{{\sst{(4)}}}
\def\5{{\sst{(5)}}}
\def\6{{\sst{(6)}}}
\def\7{{\sst{(7)}}}
\def\8{{\sst{(8)}}}
\def\n{{\sst{(n)}}}
\def\cA{{{\cal A}}}
\def\cF{{{\cal F}}}
\def\tV{\widetilde V}
\def\tW{\widetilde W}
\def\tH{\widetilde H}
\def\tE{\widetilde E}
\def\tF{\widetilde F}
\def\tA{\widetilde A}
\def\im{{{\rm i}}}
\def\tY{{{\wtd Y}}}

\def\bD{{{\bar D}}}

\newfont{\bbbold}{msbm10}
\newfont{\sbbbold}{msbm10 scaled 800}
\def\bbA{\mbox{\bbbold A}}
\def\bbB{\mbox{\bbbold B}}
\def\bbC{\mbox{\bbbold C}}
\def\bbD{\mbox{\bbbold D}}
\def\bbE{\mbox{\bbbold E}}
\def\bbF{\mbox{\bbbold F}}
\def\bbG{\mbox{\bbbold G}}
\def\bbH{\mbox{\bbbold H}}
\def\bbI{\mbox{\bbbold I}}
\def\bbJ{\mbox{\bbbold J}}
\def\bbK{\mbox{\bbbold K}}
\def\bbL{\mbox{\bbbold L}}
\def\bbM{\mbox{\bbbold M}}
\def\bbN{\mbox{\bbbold N}}
\def\bbO{\mbox{\bbbold O}}
\def\bbP{\mbox{\bbbold P}}
\def\bbQ{\mbox{\bbbold Q}}
\def\bbR{\mbox{\bbbold R}}
\def\bbS{\mbox{\bbbold S}}
\def\bbT{\mbox{\bbbold T}}
\def\bbU{\mbox{\bbbold U}}
\def\bbV{\mbox{\bbbold V}}
\def\bbW{\mbox{\bbbold W}}
\def\bbX{\mbox{\bbbold X}}
\def\bbY{\mbox{\bbbold Y}}
\def\bbX{\mbox{\bbbold X}}
\def\bbZ{\mbox{\bbbold Z}}

\def\IR{\mathbb R}
\def\ZZ{\mathbb Z}

\def\C{{\mathscr{C}}}
\def\H{{\mathcal{H}}}
\def\N{{\mathcal{N}}}
\def\M{{\mathcal{M}}}
\def\R{{\mathcal{R}}}
\def\L{{\mathcal{L}}}
\def\V{{\mathcal{V}}}
\def\W{{\mathcal{W}}}

\def\fG{{\mathfrak{G}}}
\def\fH{{\mathfrak{H}}}
\def\fP{{\mathfrak{P}}}
\def\fI{{\mathfrak{I}}}
\def\fJ{{\mathfrak{J}}}
\def\fj{{\mathfrak{j}}}

\def\Z{\mathds{Z}}

\def\e{{\boldsymbol{e}}}
\def\f{{\boldsymbol{f}}}
\def\h{{\boldsymbol{h}}}
\def\x{{\boldsymbol{x}}}
\def\y{{\boldsymbol{y}}}
\def\b{{\boldsymbol{b}}}
\def\z{{\boldsymbol{z}}}
\def\upsi{{{\boldsymbol \upsilon}}}

\newcommand{\Scal}[1]{\Bigl ({#1} \Bigr )}
\newcommand{\scal}[1]{\bigl ({#1} \bigr )}
\newcommand{\trace}{\hbox {Tr}~}

\makeindex             

\begin{document}

\title*{String Theory, Unification and Quantum Gravity\\\hfill\normalsize{\rm
Imperial/TP/12/KSS/01}
}

\author{K.S.\ Stelle}

\institute{K.S.\ Stelle \at Imperial College London, London SW7 2AZ, UK; \email{k.stelle@imperial.ac.uk}}
%
%
\maketitle

\addvspace{-2.5cm}
\textsl{\noindent
Lectures given at the 6th Aegean Summer School, ``Quantum Gravity and Quantum Cosmology'', Chora, Naxos Island, Greece, 12-17 September, 2011.}
\vskip 2cm

\abstract{An overview is given of the way in which the unification program of particle physics has evolved into the proposal of superstring theory as a prime candidate for unifying quantum gravity with the other forces and particles of nature. A key concern with quantum gravity has been the problem of ultraviolet divergences, which is naturally solved in string theory by replacing particles with spatially extended states as the fundamental excitations. String theory turns out, however, to contain many more extended-object states than just strings. Combining all this into an integrated picture, called M-theory, requires recognition of the r\^ole played by a web of nonperturbative duality symmetries suggested by the nonlinear structures of the field-theoretic supergravity limits of string theory.}

\section{Introduction: The ultraviolet problems of gravity}
\label{sec:1}

Our currently agreed picture of fundamental physics involves four principal forces: strong, weak, electromagnetic; and gravitational. The first three are well described by the Standard Model, based on the nonabelian gauge group \newline $SU(3)^{\hbox{\small strong}}\times \left(SU(2)\times U(1)\right)^{\hbox{\small electroweak}}$. In the process of unifying these forces, one necessarily had to postulate new physical phenomena going beyond the specifically desired unification. Thus, in order to make the $SU(2)\times U(1)$ electroweak unification work, one had also to accept also the neutral $Z^0$ field in addition to the desired charged $W^\pm$ intermediate vector fields (needed to resolve the interactions of the nonrenormalizable 4-fermion Fermi theory). The experimental discovery of the corresponding $Z^0$ particle was a great triumph of the Standard Model.

Another key ingredient of our current perspective is the notion of spontaneous symmetry breaking: symmetries of the field equations may be broken by the vacuum, thus becoming non-linearly realized and at the same time allowing for the generation of masses for gauge fields -- 
known as the Higgs effect. The Standard Model is moreover renormalizable: although ultraviolet infinities exist, they can be corralled into renormalizations of a finite set of parameters, thus allowing for consistent perturbative analysis of the rest of the theory. And most importantly, the Standard Model is now confirmed to very high precision by experiments at CERN, Fermilab and other laboratories.

Einstein's General Theory of Relativity, on the other hand, is nonrenormalizable, causing it to break down when interpreted as a quantum theory. One immediate indication of this is the dimensional character of the gravitational coupling constant $\kappa = \sqrt{8\pi G}$, which has dimensions of length (in units where $\hbar=c=1$). Einstein gravity's uncontrolled divergences go on to corrupt otherwise well-behaved ``matter'' theories.

Consider, for example, a radiative correction to the Higgs mass caused by a gauge-particle emission and reabsorption:

\begin{center}
\begin{fmffile}{radcorr}
\begin{fmfgraph*}(40,0)
\fmfleft{i0,d1}
\fmfright{o0,d2}
\fmf{plain,tension=1/3}{i0,i1}
\fmf{plain,tension=1/3}{o1,o0}
\fmfn{plain}{i}{2}
\fmf{plain}{i2,v,o2}
\fmfn{plain}{o}{2}
\fmffreeze
\fmf{gluon,right}{i1,o1}
\fmf{plain}{v}
\end{fmfgraph*}
\end{fmffile}
\end{center}
\vspace {1cm}

In the Standard Model, with gauge coupling constant $g$, incoming momentum $p$ and loop momentum $k$, the corresponding integral with a cutoff $\Lambda$ has the form
\be
g^2\int^\Lambda d^4k\frac{k^2}{k^2((p+k)^2+m^2)}
\ee
which has logarithmic divergences $\sim g^2\ln\Lambda\, p^2$, requiring a counterterm $(\partial\phi)^2$ and also another $\sim g^2\ln\Lambda\, m^2$, requiring a counterterm $m^2\phi^2$. Since both of these counterterm operators are present in the Standard Model Lagrangian from the start, they can be accounted for by standard wavefunction and mass renormalizations.

When the system is coupled to gravity, however, the ultraviolet divergent integrals get much worse:
\be
\kappa^2\int^\Lambda d^4k \frac{k^4}{k^2((p+k)^2+m^2)}
\ee
producing now logarithmic divergences $\sim \kappa^2\ln\Lambda\,(p^4\ ,m^2p^2\ ,m^4)$ in addition to the flat-space SM divergences. The $p^4$ divergence would require a counterterm $(\partial^2\phi)^2$, which is an operator not present in the original theory.
Moreover, this bad ultraviolet behavior gets worse and worse as the loop-order increases. At two loops, one encounters divergences $\sim \kappa^4\ln\Lambda\,p^6 + \ldots$\,, requiring a counterterm like $(\partial^3\phi)^2$\,. Each new loop adds 2 to the divergence count.
Thus, Einstein gravity is not only uncontrolled in its own divergence structure; it also renders otherwise well-behaved matter theories such as the Standard Model uncontrollable when coupled to gravity.

Pure General Relativity has a na\"{\i}ve degree of divergence at $L$ loops in spacetime dimension $D$ given by $\Delta = (D-2)L+2$. When confronting the ultraviolet problem of quantum gravity, one wants to focus on the most serious divergent structures, whose elimination would require the introduction of genuinely new operators not present in the classical Lagrangian. For this purpose, candidate counterterms that vanish subject to the classical field equations can be handled by a more standard procedure, by making field-redefinition renormalizations, which generalize the wavefunction renormalizations of renormalizable theories. Leaving these more easily handled divergence structures to one side, one searches for counterterm structures that do {\em not} vanish subject to the classical equations of motion. 

Using dimensional regularization to ensure a manifestly generally-coordinate-invariant quantization, one captures only the logarithmic divergences of a straight momentum-cutoff procedure. To balance engineering dimensions, this requires a number of factors of external momentum to be present on the external lines of a divergent diagram, in order to pick out just the logarithmically divergent part. Accordingly, at $L=2$ loops in $D=4$ dimensions, one expects $\Delta=6$, which could be achieved by counterterms like $\int d^4x\sqrt{-g}(R_{\mu\nu\rho\sigma}R^{\rho\sigma\lambda\tau}R_{\lambda\tau}{}^{\mu\nu})$ or\\ $\int d^4x\sqrt{-g} (R_{\mu\nu\rho\sigma}\square R^{\rho\sigma\mu\nu})$\, where $\square=g^{\mu\nu}\nabla_\mu\nabla_\nu$ is a covariant d'Alembertian. However, use of the Bianchi identities shows that the second of these types vanishes subject to the classical equations of motion, so it may be dealt with by field-redefinition renormalizations. Only the first is a truly dangerous type. And indeed, in pure GR, such a $(\hbox{curvature})^3$ counterterm does occur at the 2-loop order in $D=4$. \cite{Goroff:1985th, vandeVen:1991gw}

In supergravity theories, local supersymmetry places additional constraints on counterterms. This has the consequence that the 2-loop divergence of pure GR is absent. In pure supergravities, the first counterterm that does not vanish subject to the classical equations of motion (``on-shell'' in the jargon) then occurs at the 3-loop level:
\begin{center}
\includegraphics[scale=.6]{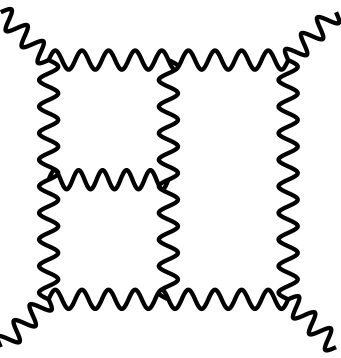}\\
\end{center}
The corresponding $D=4$ counterterm has $\Delta=8$ and starts with a purely gravitational part that is quadratic in the Bel-Robinson tensor, \ie quartic in curvatures \cite{Deser:1977nt}
\be
\int d^4x \sqrt{-g} T_{\mu\nu\rho\sigma} T^{\mu\nu\rho\sigma}\ ,\quad T_{\mu\nu\rho\sigma} = R_\mu{}^\alpha{}_\nu{}^\beta R_{\rho\alpha\sigma\beta} + {}^\ast R_\mu{}^\alpha{}_\nu{}^\beta\, {}^\ast R_{\rho\alpha\sigma\beta}
\ee
For lesser supergravities (with $N\le4$ independent gravitini), extensions of this structure remain as candidates for the first anticipated serious nonrenormalizable divergence.

\section{String theory basics}
\label{sec:2}

The fundamental excitations of String Theory are not point particles, as in ordinary quantum field theories, but extended objects. Thus, point-particle worldline interactions such as
\begin{center}
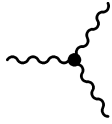

\begin{fmffile}{3grav}
\begin{fmfgraph*}(15,15)
  \fmfleft{gr1}
  \fmf{photon}{gr1,v}
  \fmf{photon}{gd,v,gu}
  \fmfright{gd,gu}
  \fmfdot{v}
\end{fmfgraph*}
\end{fmffile}\\
\captionof{figure}{3-point field-theoretic particle vertex}\label{particlevertex}
\end{center}
become smoothed out to string worldsheet interactions like 
\begin{center}
\includegraphics[scale=.5,angle=-90]{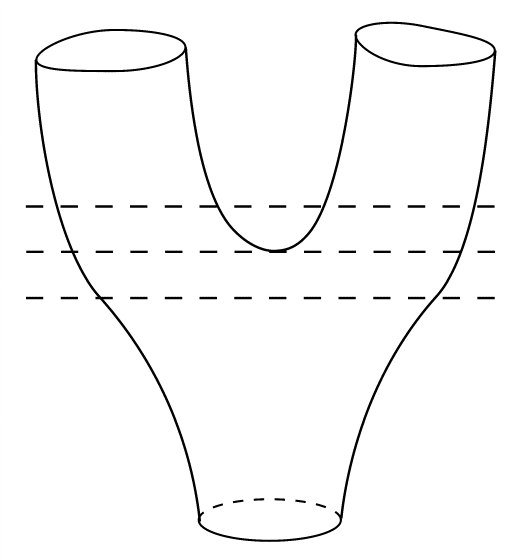}\\
\captionof{figure}{3-closed-string vertex: the splitting point is determined by the choice of time slicing, unlike the sharp identification of the interaction point in particle theory}\label{stringvertex}
\end{center}
with a consequent loss of sharpness in the spacetime localization of the interaction.

The field-theory propagator 
\begin{center}
\begin{fmffile}{prop}
\begin{fmfgraph*}(30,2)
  \fmfleft{gr1}
  \fmfright{gr2}
    \fmf{photon}{gr1,gr2}
\end{fmfgraph*}
\end{fmffile}
\strut
\end{center}
which has the usual overall momentum-space $\frac1{k^2}$ structure
becomes in closed-string theory that for a cylinder
\begin{center}
\includegraphics[scale=.38,angle=-90]{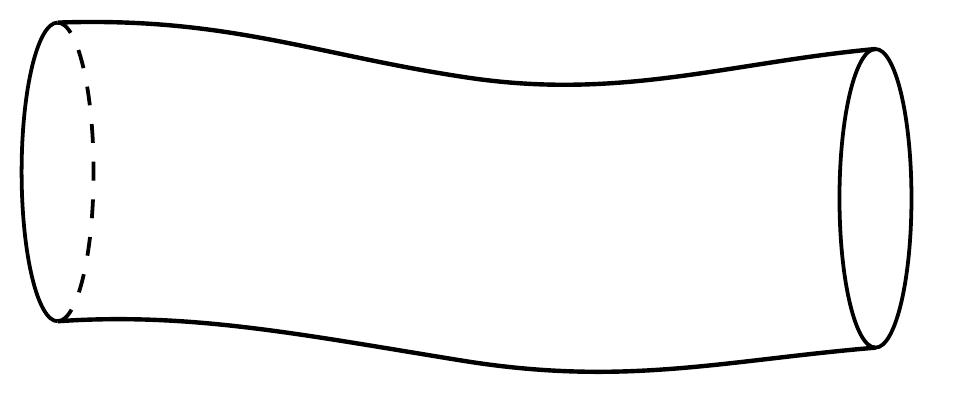}\\
\end{center}
with characteristic string length scale $\ell_s$ and momentum-space structure $\frac{e^{-\ap k^2}}{k^2}$ where $\ap$ is the {\em string slope parameter}, related to the characteristic string length scale by $\ap=\frac{\ell_s^2}{2\hbar^2c^2}$.
The decreasing exponentials arising from string propagators give rise to {\em convergent} loop diagrams for quantum corrections, yielding effectively a cutoff to the field-theory divergences at a scale $\Lambda\sim(\ell_s)^{-1}$.
\subsection{Reparametrization invariance}
\label{subsec:2.1}

An essential feature of all relativistic systems is the freedom to choose arbitrary parametrizations for their histories. Begin with the analog of a relativistic particle, whose action is obtained geometrically from the invariant proper length of its worldline as shown in Figure \ref{worldline}:
\begin{center}
\includegraphics[scale=.7]{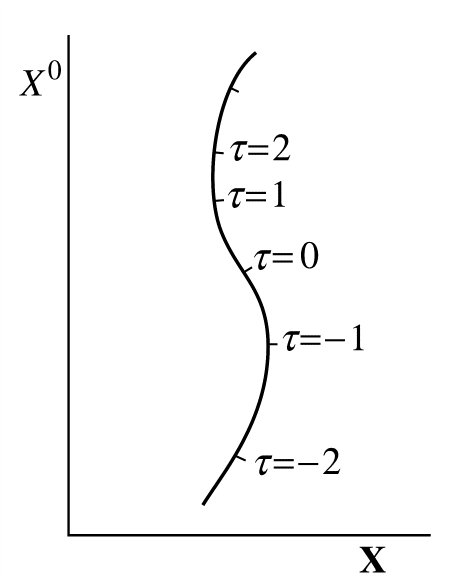}\\
\captionof{figure}{Particle worldline}\label{worldline}
\end{center}

This yields a worldline reparametrization-invariant action
\be
I_{\hbox{\scriptsize particle}}=-m \int d\tau\left(-\frac{dx^\mu}{d\tau}\frac{dx^\nu}{d\tau}g_{\mu\nu}(x)\right)^{\frac12}\ ,
\ee
which has the following manifest local invariances:
\begin{enumerate}
\item Spacetime general covariance:
\be
x^\mu \rightarrow x^{\mu'}(x)\qquad\quad g'_{\mu\nu}(x')=\frac{\partial x^\rho}{\partial x^{\mu'}}\frac{\partial x^\sigma}{\partial x^{\nu'}}g_{\rho\sigma}(x)
\ee
\item Worldline reparametrization invariance:
\be
\tau \rightarrow \tau' \qquad 
\begin{matrix}
x'^\mu(\tau')= x^\mu(\tau)&\quad\hbox{(worldline scalar)}\\
\\
\frac{dx^\mu}{d\tau}\rightarrow \frac{dx^\mu}{d\tau'}=\frac{dx^\mu}{d\tau}\frac{d\tau}{d\tau'}\quad&\quad\hbox{(worldline vector)}
\end{matrix}
\ee
\end{enumerate}

The worldline reparametrization invariance is physically important because it removes a negative-energy mode: for a metric $g_{\mu\nu}$ of Minkowski signature $(-+++\ldots)$, the $x^0(\tau)$ ``scalar field'' along the $d=1$ worldline has the wrong sign of kinetic energy. However, this potential ghost mode is precisely removed from the theory by the worldline reparametrization invariance.

As is generally the case for gauge theories, the worldline reparametrization invariance gives rise, in the Hamiltonian formalism, to a constraint on the conjugate momenta:
\be
p_\mu p_\nu g^{\mu\nu}(x)=-m^2\ , \qquad \hbox{where}\  p_\mu=\frac{\partial \L}{\partial\left(\frac{\partial x^\mu}{\partial\tau}\right)}\ .\label{momconstraint}
\ee
Thus, for a particle in $D$ dimensional spacetime, $(D-1)$ degrees of freedom remain after taking into account the worldline reparametrization invariance and the corresponding Hamiltonian constraint. The constraint \eqref{momconstraint} is recognized as the mass-shell condition for the relativistic particle.
\subsection{The string action}
\label{subsec:2.2}
Now generalize the relativistic particle action to that of a relativistic extended object with intrinsic spatial dimensionality $p=1$. Instead of a worldline, one now has a 2-dimensional worldsheet as illustrated in Figure \ref{openstring} for an open string; for a closed string, one needs to identify $\sigma=0$ and $\sigma=\pi$:
\begin{center}
\includegraphics[scale=.7]{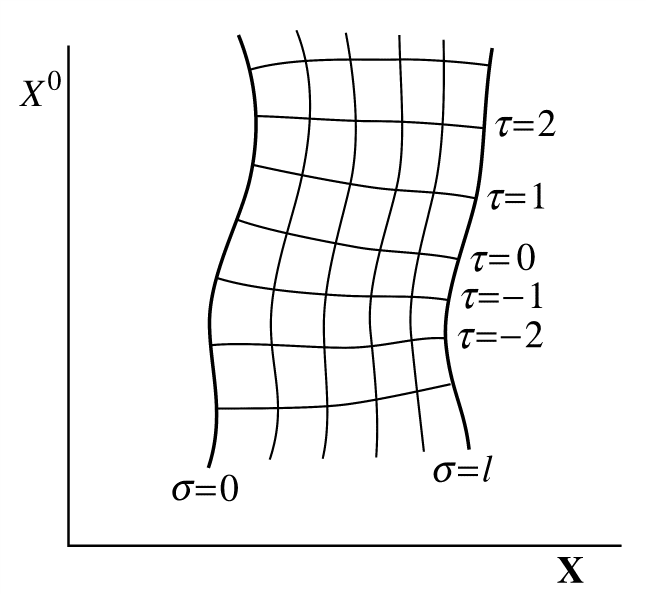}\\
\captionof{figure}{Open string worldsheet}\label{openstring}
\end{center}
The string worldsheet action is then the reparametrization-invariant area of the worldsheet $\W$
\be
I_{\hbox{\scriptsize string}}=-T\int_{\W}d^2\xi\Big(-\det\big(\partial_ix^\mu(\xi)\partial_jx^\nu(\xi)g_{\mu\nu}(x(\xi)\big)\Big)^{\frac12}\ .
\ee

As in the particle case, one has a number of local worldsheet invariances:
\begin{enumerate}
\item Spacetime general covariance $x^\mu(\tau,\sigma)\rightarrow x^{\mu'}(\tau,\sigma)$.
\item $d=2$ worldsheet reparametrization invariance ${x'}^\mu(\tau',\sigma')=x^\mu(\tau,\sigma)$.
\item Exceptionally for the $d=2 \leftrightarrow p=1$ case among the general class of ``$p$-branes'', one has an additional local worldsheet invariance: {\em Weyl invariance}.\\
Weyl invariance is crucial to the ability to carry out quantization of the string.
Jealously preserving it leads to the notion of a {\em critical dimension} for string theory.
\end{enumerate}

To see the Weyl invariance, reformulate the string action with an independent worldsheet metric $\gamma_{ij}(\xi)$ \cite{Deser:1976rb,Brink:1976sc, Polyakov:1981rd}:
\be
I_{\hbox{\scriptsize string DZBdVHP}}=-\frac12 T\int d^2\xi\sqrt{-\det \gamma}\left(\gamma^{ij}(\xi)M_{ij}\right)
\ee
where $M_{ij}=\partial_ix^\mu\partial_j x^\nu g_{\mu\nu}(x)$ is the induced metric on the worldsheet and $\gamma^{ij}$ is the matrix inverse of $\gamma_{jk}$.
Varying $\gamma_{ij}(\xi)$ as an independent field, obtain its field equation $(\gamma^{ik}\gamma^{jl}-\frac12\gamma^{ij}\gamma^{kl})M_{kl}=0$\,. Note that for $d=2$ worldsheet dimensions, the trace of this equation vanishes identically: $\gamma^{kl}M_{kl}-\frac12\gamma^{ij}\gamma_{ij}\gamma^{kl}M_{kl}\equiv0$\,.

This weakening of the set of algebraic equations for $\gamma{ij}$ corresponds to the local Weyl invariance of the DZBdVHP action:
\bea
\gamma_{ij} &\rightarrow& \Omega(\xi)\gamma_{ij}\\
\xi^i &\rightarrow& \xi^i
\eea
where $\Omega(\xi)$ is an arbitrary positive local scale factor. Varying the string action, one obtains the algebraic equation determining $\gamma_{ij}(\xi)=\Omega(\xi)M_{ij}$, with $\Omega(\xi)$ left undetermined, and the $d=2$ covariant wave equation for $x^\mu(\xi)$: 
\be
\nabla^i_{(\gamma,g)}\partial_ix^\mu=0\,.\label{waveeqn}
\ee

For closed bosonic strings, the wave equation \eqref{waveeqn}, plus periodicity in the spatial worldsheet coordinate $\sigma$ (conventionally taken to identify $\sigma=0$ with $\sigma=\pi$), give the full classical dynamical system of closed-string equations. 

For open strings, the $\sigma$ coordinate is conventionally considered to take its values in the closed interval $\sigma\in [0,\pi]$.
Then, considering also the surface term arising in the variation of $I_{\hbox{\scriptsize DZBdVHP}}$ upon integration by parts, one finds in addition the following Neumann boundary conditions:
\be
M_{0i}\epsilon^{ik}\partial_k x^\mu=0\qquad \hbox{at}\ \ \sigma=0\,,\ \pi\ .
\ee

Considering strings in a flat spacetime background, $g_{\mu\nu}=\eta_{\mu\nu}$\,, and picking conformal gauge for the worldsheet reparametrization symmetries, $\gamma_{ij}=\Omega(\xi)\hbox{diag}(-1,1)$\,, the $x^\mu(\xi)$ wave equation and open-string boundary conditions become
\bea
\square x^\mu &=& 0\quad \hbox{where $\square=\eta^{ij}\partial_i\partial_j$ is the flat-space $d=2$ d'Alembertian}\\
\frac\partial{\partial\sigma} x^\mu &=& 0\quad \hbox{at}\ \ \sigma=0\,,\ \pi \ .
\eea
These may be interpreted classically as requiring waves to travel back and forth along the string at speed $c=1$, while the boundary conditions imply that the endpoints of the open string travel through the embedding spacetime at speed $c=1$.

For closed strings, there are periodicity conditions instead of reflective boundary conditions. In that case, there can be independent left- and right-moving waves travelling around the string at speed $c=1$.

A simple solution to the open-string equations of motion and boundary conditions is
\ba
x^0 &=\frac12(p^+ + \frac{A^2}{p^+})\tau& x^3 &=\frac12(p^+-\frac{A^2}{p^+})\tau\\
x^1 &= A\cos\sigma \cos\tau & x^2 &= A\cos\sigma \sin\tau\ .
\ea
Boosting to a Minkowski reference frame where $x^3=0$, find $p^+=\frac{A^2}{p^+}=\pm A$; in this frame, the center-of-mass of the open string at $\sigma=\pi/2$ remains stationary while the string profile at any time $\tau$ describes a straight line of length $2A$ rotating with period $2\pi A$ (with respect to the background Minkowski time $t=x^0=A\tau$).

The total string energy for this solution is $E=\frac\pi2\ell T$, where $\ell=2A$ is the string length. Thus, the parameter $T$ should be interpreted as the string tension.

The angular momentum for this solution is $J^3=\frac\pi8 \ell^2 T=\frac{E^2}{2\pi T}$. This linear relationship between angular momentum and $(\hbox{energy})^2$ is known as {\em Regge behavior}.

One can now make a rough Bohr-Sommerfeld estimate of the quantum spectrum, requiring $|J|=n\hbar\ ,\ n\in \Z$ and considering the excitations in their rest frames where $E=M$. Then $n=\frac{|J|}\hbar=\ap M^2$ where $\ap=\frac1{2\pi\hbar cT}$ is the string slope parameter. The quantized states lie on linear Regge trajectories making an angle $\ap$ in a $J/\hbar$ versus $M^2$ plot:
\begin{center}
\includegraphics[scale=.7]{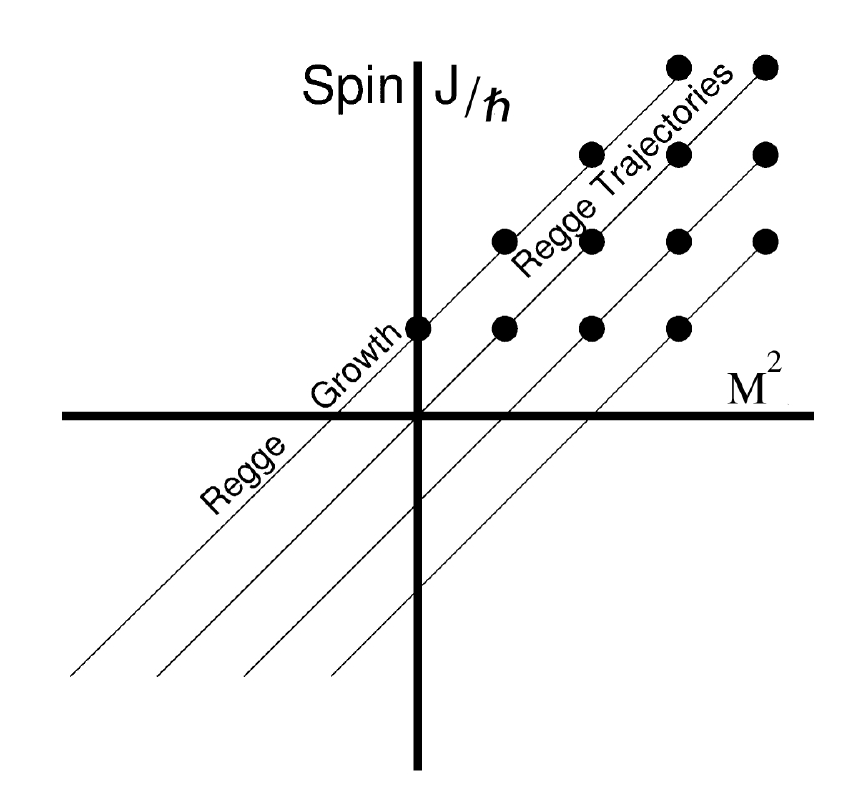}\\
\captionof{figure}{Linear Regge trajectories relating spin and $({\rm mass})^2$ of particle states}
\end{center}
Of course, finding such Regge trajectories in the physical particle spectrum would be a spectacular confirmation of string theory.

The above semiclassical analysis makes the lowest-lying string state a massless scalar. However, a more careful quantum analysis reveals a feature missed by the Bohr-Sommerfeld analysis: the intercept at $n=0$ is shifted down: $\ap E^2=n-1 \leftrightarrow M^2=\frac{n-1}{\ap c^4}$.
Thus, the $n=0$ lowest-lying state of the bosonic string becomes a negative $M^2$ tachyon, while the $n=1$ first excited state with $|J|=\hbar$ becomes massless. Accordingly, the open-string quantum spectrum contains massless spin 1 gauge fields.

The closed string dispenses with the reflective open-string boundary conditions and accordingly has twice as many modes: independent left- and right-moving excitations. It turns out that the closed-string spectrum is a tensor product of open-string spectra in the R \& L sectors, together with a level-matching condition: the R and L level numbers must be equal.
The closed-string $(n_L,n_R)=(1,1)$ states thus contain the tensor product of $(\hbox{spin}\, 1)_L\times(\hbox{spin}\, 1)_R$ states: the closed-string spectrum contains {\em massless spin 2}.

\section{Effective field equations}
\label{sec:3}

The spin 2 mode identified in the closed-string spectrum is not merely a hint that closed-string theory has something to do with gravity. The full Einstein action also emerges when one considers string theory from an effective-field-theory point of view. The key to understanding this is the requirement that anomalies in the local Weyl symmetry cancel. 

Analysis of the spectrum of any string theory shows the presence of at least three types of massless field: the graviton $g_{\mu\nu}(x)$, a 2-form antisymmetric tensor gauge field $B_{\mu\nu}(x)$ and a ``dilatonic'' scalar $\phi(x)$. In supersymmetric theories, the infamous tachyon of bosonic string theory is absent. In non-supersymmetric contexts, the tachyon is interpreted as indicating that the presumed ``vacuum'' around which one is trying to quantize is unstable and so one should shift instead to a stable vacuum background. This shift is made explicit in string field theory.

To begin with, consider just the massless backgrounds $(g_{\mu\nu}(x), B_{\mu\nu}(x), \phi(x))$\,.
The string action on this effective-field background is then
\ba
I_{\hbox{\scriptsize gen.\ back.}}& =\nn\\
-\frac1{4\pi\ap}&\int d^2\xi\sqrt{-\gamma}\Big[(\gamma^{ij}g_{\mu\nu}(x) - \epsilon^{ij}B_{\mu\nu}(x))\partial_i x^\mu \partial_j x^\nu + \ap R(\gamma)\phi(x)\Big]
\ .
\ea

Note that the 2-form background gauge field $B_{\mu\nu}(x)$ has rank needed to pull back using $\partial_ix^\mu$ to a 2-form on the worldsheet, precisely as needed to contract with the $d=2$ Levi-Civita tensor $\epsilon^{ij}$. Note also that the coupling to the dilaton $\phi(x)$ involves the worldsheet Ricci scalar $R(\gamma)$ and enters with an additional factor of $\ap$, as is appropriate if $g_{\mu\nu}$, $B_{\mu\nu}$, $\phi$ and $\gamma_{ij}$ are all taken to be dimensionless.

The worldsheet Weyl symmetry $\gamma_{ij}(\xi)\rightarrow \Omega(\xi)\gamma_{ij}(\xi)$ is respected by the $B_{\mu\nu}$ coupling (since $\sqrt{-\gamma}\epsilon^{ij}_{\hbox{\tiny tensor}}=\epsilon^{ij}_{\hbox{\tiny density}}$ is $\gamma_{ij}$ independent), but it is violated by the dilaton coupling $\phi R(\gamma)$. This is intentional: the dilaton coupling is introduced precisely to complete the cancellation of Weyl-symmetry anomalies arising in the perturbative $\ap$ expansion.

$I_{\hbox{\scriptsize gen.\ back.}}$ is manifestly invariant under spacetime general coordinate transformations $x^\mu \rightarrow x^{\mu'}$ provided $g_{\mu\nu}$ and $B_{\mu\nu}$ transform as tensors and the dilaton $\phi$ is a scalar. It is also invariant under the $B_{\mu\nu}$ gauge transformation $B_{\mu\nu}\rightarrow B_{\mu\nu}+\partial_\mu\zeta_\nu-\partial_\nu\zeta_\mu$, which causes the integrand of $I_{\hbox{\scriptsize gen.\ back.}}$ to vary by a total derivative.

The general-coordinate and 2-form gauge invariances are precisely what are needed to give agreement with the expected degree-of-freedom counts for these massless backgrounds:
$
\begin{matrix}\large(\frac12 D(D-3)\ ,\ \ \\ \hbox{\scriptsize metric}\end{matrix} \begin{matrix}\frac12 (D-2)(D-3)\large)\\ \hbox{\scriptsize 2-form}\end{matrix}\ .
$

Imposing on this background-coupled string system the requirement that the Weyl symmetry anomalies cancel gives differential-equation restrictions on the background fields ($g_{\mu\nu}$, $B_{\mu\nu}$, $\phi$); these may be viewed as {\em effective field equations} for these massless modes. 

The system of effective field equations for ($g_{\mu\nu}$, $B_{\mu\nu}$, $\phi$) is, remarkably, derivable from an {\em effective action} for the $D$ dimensional massless modes \cite{Callan:1985ia}.
\bea
I_{\hbox{\scriptsize eff}}&=&\int d^Dx\sqrt{-g}e^{-2\phi}\Big[(D-26)-\frac32\ap(R+4\nabla^2\phi
-4(\nabla\phi)^2\nn\\
&&\hspace{5cm} -\ft1{12}F_{\mu\nu\rho}F^{\mu\nu\rho}+{\cal O}(\ap)^2 \Big]\ .
\eea

Note the appearance of a {\em critical dimension}\,: the ``cosmological term'' vanishes only for $D=26$, showing that, for a flat background, the Weyl anomalies can be cancelled in this way only in 26 dimensional spacetime. 

In superstring theories, there are additional anomaly contributions from the fermionic modes which change the critical dimension to 10. Moreover, in supergravity theories, the tachyon is absent, so $D=10$ flat space becomes a stable background of the massless modes. Aside from the change of the critical dimension to 10, however, the above effective action remains valid for a subset of the bosonic background of the theory, known as the Neveu-Schwarz sector.

Now specialize to $D=10$ for the superstring and accordingly drop the cosmological term. Moreover, the unfamiliar $e^{-2\phi}$ factor in front of the Ricci scalar $R$ may be eliminated together with the $4e^{-2\phi}\nabla^2\phi$ term by redefining the metric: 
$g^{\rm (e)}_{\mu\nu}=e^{-\phi/2}g^{\rm (s)}_{\mu\nu}$ where $g^{\rm (s)}$ is the previous string-frame metric and $g^{\rm (e)}$ is the new Einstein-frame metric.

In the Einstein frame, the Neveu-Schwarz sector effective action then becomes
\be
I_{\hbox{\scriptsize Einstein}}=\int d^{10}x\sqrt{-g^{\rm (e)}}\Big[R(g^{\rm (e)}) -
\ft12\nabla_{\mu}
\phi\nabla^{\mu}\phi-\ft1{12}e^{-\phi}F_{\mu\nu\rho}F^{\mu\nu\rho}\Big]\ .
\ee

Including effective-action contributions for the other (Ramond sector) bosonic backgrounds and also for fermionic backgrounds, one obtains thus a correspondence between superstring theories and related supergravity theories: a supergravity theory describes the massless field-theory sector of the corresponding superstring theory. One obtains in this way effective supergravity theories for the following superstring theory variants: type IIA, type IIB, type I with gauge group ${\rm SO}(32)$, heterotic ${\rm SO}(32)$ and heterotic $E_8\times E_8$.

\section{Dimensional reduction and T-duality}
\label{sec:4}

In order to extract a more realistic physical scenario from the higher-dimensional contexts native to string theory, one needs to reduce the effective theory down to $D=4$ one way or another. The most straightforward way to do this is by a traditional Kaluza-Klein reduction. 

The basic idea can be explained in terms of a massless scalar field in $D=5$ on a spacetime with the $5^{\rm th}$ direction periodically identified: $y \sim y + 2\pi \R$. Periodicity requirements on the de Broglie waves $e^{ipy/\R}$ then require the momenta in the $y$ direction to be quantized, $p_n=\frac{n\hbar}{\R}$. Thus, expand the $D=5$ field $\phi(x^\mu,y)$, $\mu=0,1,2,3$, using a complete set of eigenfunctions of the Laplace operator on a circle, \ie in terms of plane waves with quantized momenta:
\be
\phi(x^\mu,y)=\sum_{n\in\Z}\phi_n(x^\mu)e^{iny/\R}\ .
\ee

Inserting this expansion into the $D=5$ Klein-Gordon field equation gives an infinite number of $D=4$ equations for the independent modes $\phi_n(x^\mu)$:
\be
\frac1{c^2}\frac{\partial^2\phi_n}{\partial t^2}-\nabla^2\phi_n+\frac{n^2}{\R^2}\phi_n=0\ .
\ee
Thus, the $n\ne0$ modes $\phi_n$ are massive, with masses $m_n=\frac{n}{\R}$.

The basic physical picture is that at energies low compared to $\frac\hbar{c\R}$, the massive modes $\phi_{n>0}$ are frozen out, so the theory effectively reduces to just $\phi_0$.
Dimensional reduction of the supergravity theories associated to the various $D=10$ string theories produces the family of supergravity theories existing in lower spacetime dimensions, including the maximally extended $N=8$ supergravity in $D=4$.

\subsection{Dimensional reduction of strings and T-duality}
\label{subsec:4.1}

Consider now string theory in a background spacetime with a compactified direction, $x^{\sst M}\rightarrow (x^\mu,y)$, $\mu=0,\ldots,(D-2)$. The Regge towers of string states can be individually treated as particle fields; massless string states give rise to massless states in the $(D-1)$ lower dimensions plus Kaluza-Klein towers of states with masses  $\frac{n}{\R}$, just like in Kaluza-Klein field theory.

Strings, however, can do something different from particles in that they can {\em wrap around} the compactified dimension. Consider a closed-string mode expansion
\ba
x^{\sst M}(\tau,\sigma)=q^{\sst M}(\tau) &+ p^{\sst M}\ell^2\tau + 2\tilde n\R\sigma\delta^{\sst M}_y\nn\\
& + \frac{i\ell}2\sum_{k\ne0}\left(\frac{\alpha^{\sst M}_k}k e^{-2ik(\tau-\sigma)} + \frac{\tilde\alpha^{\sst M}_k}k e^{-2ik(\tau+\sigma)}\right)
\ea
where $n,\tilde n\in\Z$ and $\ell^2=2\ap$, the $(\hbox{string length})^2$.

As expected, the momentum in the compactified direction is quantized, $p^y=\frac{n}{\R}$.
However, owing to the fact that the string can wind around the compactified $y$ dimension a number $\tilde n$ times, the energy (\ie mass) formula for the string spectrum considered from the viewpoint of the dimensionally reduced theory has a generalized form:
\be
M^2=\frac{\hbar^2}{c^2}(\frac{n^2}{\R^2} + \frac{\tilde n^2\R^2}{{\ap}^2}) + \hbox{contributions from ordinary oscillator modes}\ .
\ee

This mass formula suggests a striking symmetry of string theory that is not present for particle theories: interchanging $n\leftrightarrow \tilde n$ and simultaneously inverting the compactification radius, $\R\rightarrow \ap/{\R}$ leaves the spectrum invariant. 

This symmetry is {\em T-duality:} a string propagating on a compact direction of radius $\R$ with momentum mode $n$ and winding mode $\tilde n$ is equivalent to a string propagating on a compact direction of radius ${\ap}/{\R}$ with interchanged mode numbers: momentum $\tilde n$ and winding $n$.
\begin{center}
\includegraphics[scale=.8]{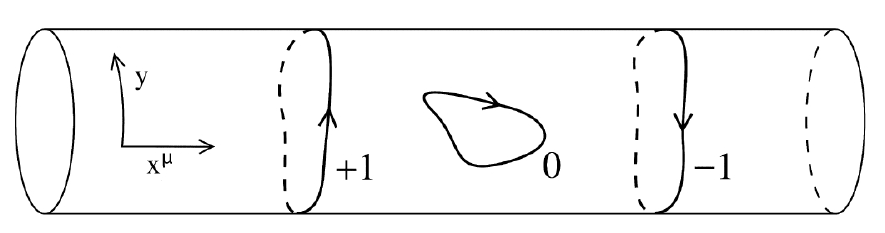}\\
\captionof{figure}{Winding modes with various $\tilde n$ values}
\end{center}

Because string and background configurations related by a T-duality transformation are {\em identified}, this symmetry, although discrete, extends the notion of local symmetry in string theory beyond the ordinary context of general coordinate and gauge invariances.

T-duality has a dramatic effect on curved background geometries. Start from a simplified closed-string action without the dilaton:
\be
I_{g,B} =-\frac12\int d^2\xi\sqrt{-\gamma}(\gamma^{ij}\partial_ix^{\sst M}\partial_jx^{\sst N} g_{\sst{MN}} - \epsilon^{ij}\partial_ix^{\sst M}\partial_jx^{\sst N} B_{\sst{MN}})\ .
\ee
Now suppose that there is an {\em isometry} in the $y$ direction, \ie that $g_{\sst{MN}}$ and $B_{\sst{MN}}$ don't depend on $y$. Of course, $y(\tau,\sigma)$ is still a string variable -- the string is not prevented from moving in the $y$ direction of spacetime. But the background functional dependence on $y$ is trivial owing to the isometry. Accordingly, the string variable $y(\tau,\sigma)$ appears only through its derivative $\partial_iy$.

Now replace $\partial_i y$ everywhere in the action by $v_i$, a worldsheet vector. Enforce the curl-free nature of $v_i$ by a Lagrange multiplier term $\int d^2\xi \sqrt{-\gamma}\epsilon^{ij}\partial_i z v_j$. Then eliminate $v_i$ by its algebraic equation of motion. The result is the T-dualized version of the string action written in terms of $\tilde x^{\sst {\tilde M}}(\tau,\sigma)=(x^\mu(\tau,\sigma),z(\tau,\sigma))$.

The net effect of a T-duality transformation may be seen by reassembling the results into an action $I_{\tilde g,\tilde B}$ of the same general form as $I_{g,B}$ but now for string variables $\tilde x^{\sst{\tilde M}}(\tau,\sigma)$  and with dualized backgrounds $\tilde g_{\sst{\tilde M\tilde N}}(\tilde x)$, $\tilde B_{\sst{\tilde M\tilde N}}(\tilde x)$ given by \cite{Buscher:1987sk}
\ba
\tilde g_{\mu\nu}&=g_{\mu\nu} + g^{-1}_{yy}(B_{\mu y}B_{\nu y}-g_{\mu y} g_{\nu y})\nn\\
\tilde g_{\mu z}&= g^{-1}_{yy} B_{\mu y}\qquad\qquad \tilde g_{zz}=g^{-1}_{yy}\nn\\
\tilde B_{\mu\nu}&=B_{\mu\nu} + g^{-1}_{yy}(g_{\mu y}B_{\nu y}-g_{\nu y}B_{\mu y})\nn\\
\tilde B_{\mu z}&=g^{-1}_{yy}g_{\mu y}\nn
\ea

Careful attention to the effect of T-duality transformations reveals that they can map not only between different solutions of a given string theory, but they can even map between solutions of {\em different} string theories. In particular, paying careful attention to the effect on spinor backgrounds shows \cite{Dai:1989ua, Dine:1989vu, Cvetic:1999zs} that T-duality maps between type IIA and type IIB closed-string theories: 
\be
\hbox{Type IIA on $S^1$ of radius $\R$}\quad \atop{T}{\longleftrightarrow} \quad \hbox{Type IIB on $S^1$ of radius ${\ap}/{\R}$}\nn
\ee

\section{M-Theory and the Web of Dualities}
\label{sec:5}

Another essential duality symmetry of string theory is strong-weak coupling duality, or S-duality. The {\em dilaton} field plays a crucial r\^ole in this, as its expectation value serves as the {\em coupling constant} for string interactions. String theory has no other \`a priori determined parameters (except for the scale-setting slope parameter $\ap$). 

All the essential coupling constants are determined by vacuum expectation values of scalar fields present in the theory, with coupling constants typically given by the  VEVs of exponentials like $e^\phi$. Since, in a dimensional-reduction context, massless scalar fields derive from the moduli of the reduction manifold (\eg torus circumferences, twist parameters, {\it etc.}), scalar fields with undetermined vacuum expectation values are generically called {\em moduli} fields.

The most accessible illustration of the geometry of such moduli and the symmetries acting upon them is to be found in the massless sector of Type IIB theory, whose effective action is Type IIB supergravity. The bosonic part of the action for Type IIB supergravity is
\ba
I^{{\rm IIB}}_{10} =\int d^{10}x\Bigl[e\R &+\ft14 e\, {\rm tr}(\nabla_\mu{\cal M}^{-1}\,\nabla^\mu{\cal M}) -\ft1{12}e\, H_{[3]}^T\,{\cal M}\, H_{[3]}\nn\\ 
& -\ft1{240} e\, H_{[5]}^2 - \ft1{2\sqrt2}
\epsilon_{ij}{\,}^\ast (B_{[4]}\wedge  dA_{[2]}^{(i)}\wedge dA_{[2]}^{(j)})\Bigr]\ ,
\ea
subject to the further constraint of self-duality for the 5-form field strength $H_{\mu_1\ldots\mu_5}=\frac1{5!}\epsilon_{\mu_1\ldots\mu_5\mu_6\ldots\mu_{10}}H^{\mu_1\ldots\mu_{10}}$\,. The 3-form field strengths $\sz{H_{[3]}=\begin{pmatrix}dB^1_{[2]}\\dB^2_{[2]}\end{pmatrix}}$ contract into the $2\times2$ matrix built from the scalars $\phi$ and $\chi$
\be
\M=\begin{pmatrix}e^{-\phi}+\chi^2e^\phi &\chi e^\phi\\ \chi e^\phi &e^\phi\end{pmatrix}\ .
\ee
Multiplying out the scalar kinetic terms, one finds a more familiar form:
\be
-\frac12\int d^{10}x \sqrt{-g}(\partial_\mu\phi\partial_\nu\phi g^{\mu\nu}+e^{2\phi}\partial_\mu\chi\partial_\nu\chi g^{\mu\nu})\ .
\ee

From the above form of the IIB action, one can see that it has an $SL(2,\bbR)$ symmetry $\M\rightarrow \Lambda\M\Lambda^T\ ,\ H_{[3]}\rightarrow(\Lambda^T)^{-1}H_{[3]}\ ,\ H_{[5]}\rightarrow H_{[5]}$ where $\Lambda=\begin{pmatrix}a&b\\c&d\end{pmatrix}$ with $\det\Lambda=1$ is an ${\rm SL}(2,\bbR)$ matrix. While the action of ${\rm SL}(2,\bbR)$ on $\M$ is linear, the action on $(\phi,\chi)$ is nonlinear: these fields form an ${\rm SL}(2,\bbR)/U(1)$ nonlinear sigma model. The action of ${\rm SL}(2,\bbR)$ on the scalars may be reformulated in terms of its action on the modular field $\tau=\chi+ie^{-\phi}$, which transforms in a fractional linear fashion as $\tau\rightarrow \frac{a\tau+b}{c\tau+d}$.

At the nonperturbative quantum level, the ${\rm SL}(2,\bbR)$ symmetry gets reduced to its discrete subgroup ${\rm SL}(2,\bbZ)$. This is necessary in order for the Gauss's law charges associated to $H^i_{[3]}$ to obey a Dirac quantization condition; ${\rm SL}(2,\bbZ)$ is the subgroup that preserves the resulting charge lattice. The surviving ${\rm SL}(2,\bbZ)$ may be considered to be generated by two elementary transformations, $\tau\rightarrow \tau+1$ and  $\tau\rightarrow -\frac1\tau$. For $\chi=0$, the second of these inverts the v.e.v.\ of $e^\phi$, hence the string coupling constant $g_s$. So this is called {\em S-duality} because it exchanges strong and weak string coupling.

Given that apparently different string theories can be related by T-duality transformations and that different coupling-constant regimes can be related by S-duality transformations, one naturally searches for the full interrelated set of theories and coupling regimes related by duality transformations, known as the ``web of dualities''.

A key link in this web of dualities concerns the strong-coupling limit of type IIA theory. There is no known duality that gives this limit purely within the type IIA theory, but the relation between string-theory dualities and supergravity dualities does suggest what the strong-coupling regime of type IIA string theory might become. There is one more maximal supergravity theory which had not yet been integrated into the general picture of string \& supergravity theories: {\em supergravity in 11-dimensional spacetime}. This theory has as bosonic fields just the metric $g_{\sst{MN}}$ and a 3-form gauge field $C_{\sst{MNP}}$, and as fermionic field the gravitino $\psi^\alpha_{\sst M}$ ($\alpha=1,\ldots,32$). Overall there are 128 bosonic and 128 fermionic physical degrees of freedom per spacetime point.

$D=11$ supergravity contains no scalar fields, but when it is dimensionally reduced to $D=10$ on a circle $S^1$, straightforward Kaluza-Klein reduction generates one scalar, basically from the $g_{\sst{11\,11}}$ component of the $D=11$ metric. The reduced theory precisely reproduces $D=10$ type IIA theory at the classical level, with the Kaluza-Klein scalar $\phi$ becoming the dilaton of the type IIA theory and $g_s=\large\langle e^{\phi}\large\rangle$ being the supergravity realization of the type IIA string coupling constant. Since $g_{\sst{11\,11}}$ gives the metric on the reduction circle $S^1$, the modulus field $\phi$ controls the circumference of that circle. Thus, strong coupling, $g_s\rightarrow\infty$, corresponds to the limit where the $S^1$ reduction circle circumference tends to infinity.

Now consider just {\em compactification} of $D=11$ supergravity instead of {\em dimensional reduction} down to $D=10$, \ie define the theory on a circle $S^1$ but don't discard the Kaluza-Klein towers of massive states. Taking the limit $g_s\rightarrow\infty$ now corresponds to returning the theory to uncompactified $D=11$ supergravity.

If there is to be a $D=11$ picture of $D=10$ Type IIA theory, where can the Kaluza-Klein towers of states come from? Well, the dimensional reduction of massless $D=11$ states produces massive states that also carry a $U(1)$ charge corresponding to the Kaluza-Klein vector, derived from $g_{\sst{11},\mu}$: they are $\half$ BPS states originating in the Ramond sector of the theory. And, in fact, Type IIA theory does have just such states: the tower of $\half$ BPS {\em black hole} states, carrying charges under the vector gauge field $A_\mu$ of the Type IIA theory.\,\cite{Hull:1994ys, Witten:1995ex}

For increasing $g_s=\large\langle e^{\phi}\large\rangle$, the spacing between the BPS mass levels decreases, approaching a continuum as one approaches the decompactification limit of infinite $S^1$ circumference, where the full $D=11$ nature of the theory becomes more and more manifest.
Accordingly, the strong $g_s$ coupling limit of Type IIA string theory is hypothesized to be described by a phase whose full quantum properties remain incompletely known, but which has $D=11$ supergravity as a field-theory limit. This phase of the overall picture has been called {\em M-Theory}.

\begin{center}
\includegraphics[scale=.75]{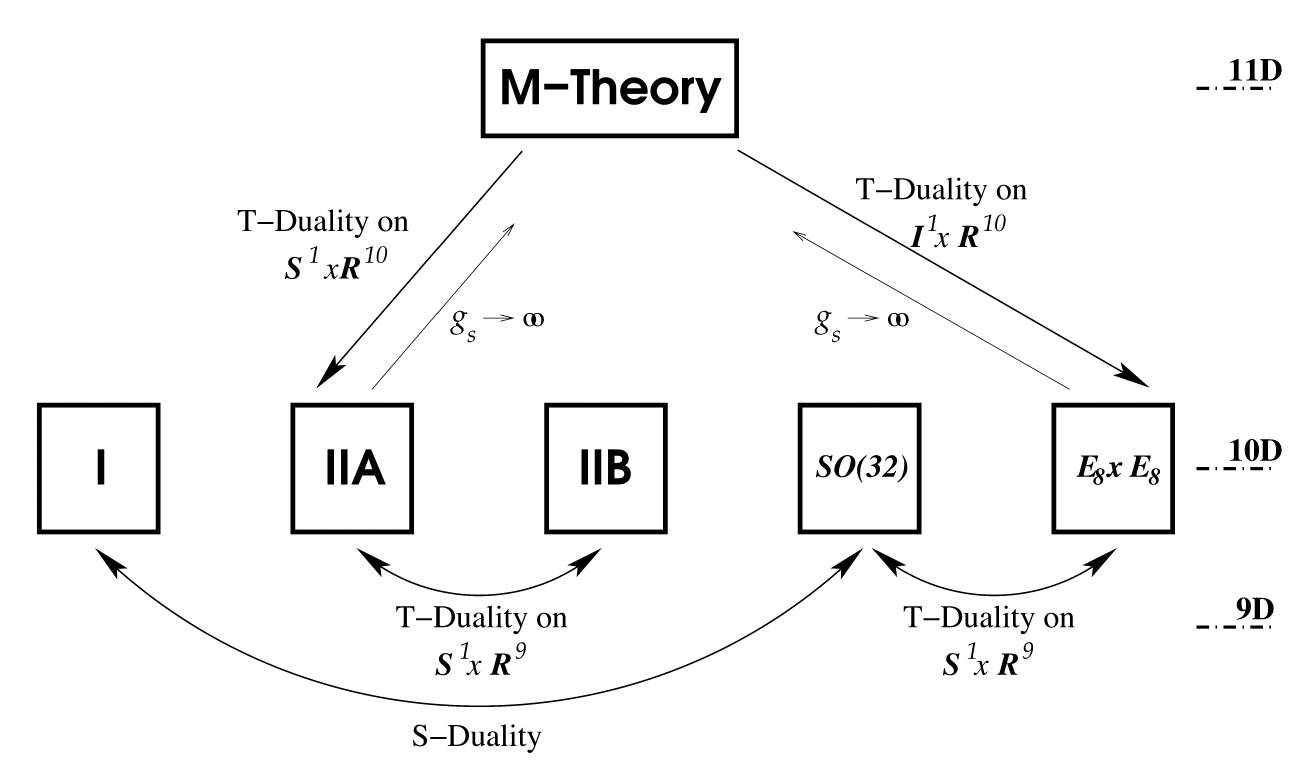}
\captionof{figure}{``Not to take this web of dualities as a sign that we are on the right track would be a bit like believing that God had put fossils into the rocks in order to mislead Darwin about the evolution of life.''  Steven Hawking}
\end{center}

\section{Branes and duality}
\label{sec:6}

Consider a T-duality transformation on the worldsheet variable $y(\tau,\sigma)$ now in a bit more detail, specializing to a flat background spacetime in which $g^{yy}=k$. The relevant part of the string action is
$-\frac k2\int d^2\xi\sqrt{-\gamma}\gamma^{ij}\partial_iy\partial_jy$\,.
Now replace $\partial_iy \rightarrow v_i$ and include as before a Lagrange multiplier $z(\tau,\sigma)$ in order to enforce the vanishing of $\partial_iv_j-\partial_jv_i$\,:  $\int d^2\xi(-\frac k2\sqrt{-\gamma}\gamma^{ij}v_iv_j + \epsilon^{ij}v_i\partial_jz)$. For $v^i$, find the algebraic equation $v^i=\frac1k\epsilon^{ij}\partial_jz$. Substituting back into the action then gives the T-dualized result $-\frac1{2k}\int d^2\xi\sqrt{-\gamma}\gamma^{ij}\partial_iz\partial_jz$\,.

Now consider, however, the effect of the above procedure on the usual open-string Neumann boundary condition $\partial_\sigma y=0$ at the endpoints (endpoint worldline normal derivative vanishes). After T-dualization, this becomes $\partial_\tau z=0$ at the endpoints (endpoint worldline tangential derivative vanishes). Thus, for the T-dualized coordinate $z(\tau,\sigma)$, one obtains a {\em Dirichlet boundary condition} $z=\hbox{constant}$ at the endpoints:
\be
\hbox{Neumann b.c.}\quad \atop{T}{\longleftrightarrow} \quad \hbox{Dirichlet b.c.}\nn
\ee

The surfaces on which Dirichlet boundary conditions are imposed obviously would break Lorentz invariance if they were considered to be imposed externally to the theory. However, considering them to be dynamical objects similar to solitons in the theory restores Lorentz symmetry. 
\begin{center}
\includegraphics[scale=.8]{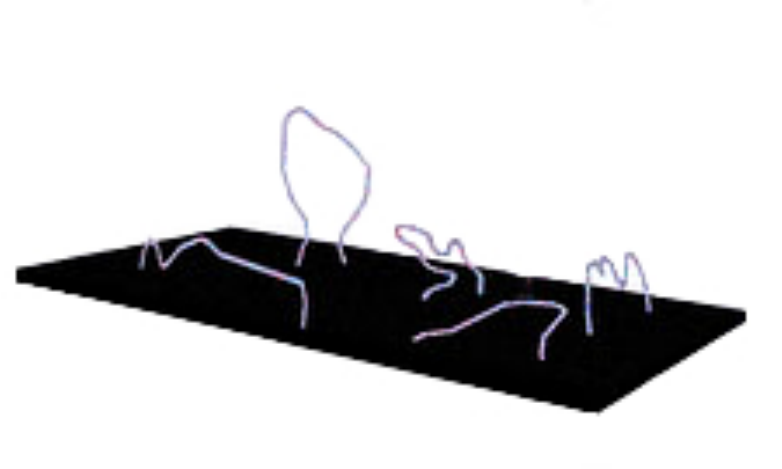}\\
\captionof{figure}{Open strings starting and ending with Dirichlet boundary conditions on a $p$-dimensional D-brane hyperplane in the target spacetime}
\end{center}

Analysis of the open-string modes in which $p$ background spatial dimensions are treated with Dirichlet and the remaining $(10-(p+1))$ spatial dimensions with Neumann boundary conditions reveals modes associated to the $(p+1)$ dimensional ``worldvolume'' of the Dirichlet surface ($p$ spatial dimensions plus time). These are a massless $U(1)$ gauge field $A_i$ together with $(9-p)$ massless scalar modes.

The massless worldvolume modes can be interpreted as Goldstone modes for the broken antisymmetric tensor gauge symmetry and for the Poincar\'e translation symmetries broken by the choice of Dirichlet boundary-condition integration constants. In other words, these massless scalar worldvolume modes may be seen to describe motions of the Dirichlet surface transverse to the worldvolume. This dynamical object is called a ${\rm D}_p$ brane.
The motions of a ${\rm D}_p$ brane are described by an effective action of Dirac-Born-Infeld type \cite{Fradkin:1985ys}:
\ba
I_{{\rm D}_p}&=-T_p\int d^{p+1}\xi e^{-\phi(x(\xi))}\left[-\det(M_{ij}+B_{ij} + 2\pi\ap F_{ij})\right]^{\frac12}\nn\\
M_{ij}&=\partial_ix^\mu\partial_jx^\nu g_{\mu\nu}(x)\quad B_{ij}=\partial_ix^\mu\partial_jx^\nu B_{\mu\nu}(x)\quad F_{ij}=\partial_iA_j(\xi)-\partial_jA_i(\xi)\ .
\ea

The dynamical extended-object hypersurfaces encountered as ${\rm D}_p$-branes in string theory have natural analogue $p$-brane solutions in the associated supergravity theories. In fact, the supergravity solutions extend the brane family beyond those seen directly as ${\rm D}_p$ branes in perturbative string theory, indicating a yet richer family of nonperturbative extended-object solutions.

A representative example is the string itself, viewed now as an extended-object solution to the effective theory's field equations. In the various $D=10$ supergravities associated to superstring theories, one always has a Neveu-Schwarz sector
\be
I_{\rm NS}=\int d^{10}x\sqrt{-g}\left[R-\frac12\nabla_\mu\phi\nabla^\mu\phi-\frac1{12}H_{\mu\nu\rho}H^{\mu\nu\rho}\right]
\ee
where $H_{\mu\nu\rho}=\partial_\mu B_{\nu\rho}+\partial_\nu B_{\rho\mu}+\partial_\rho B_{\nu\mu}$.

This effective action has an explicit solution:
\ba
ds^2 &= \H^{-\frac34}(y) dx^idx^j\eta_{ij} + \H^{\frac14}(y)dy^mdy^m\nn\\
B_{ij} &=\epsilon_{ij}\H^{-1}(y)\nn\\
e^\phi &= \H^{-\frac12}(y)\qquad \H(y)=1 + \frac{k}{(y^my^m)^3}\ .
\ea
The singular surface at $y=\sqrt{y^my^m}=0$, parametrized by $x^i$, $i=0,1$, corresponds to the static worldsheet of an infinite string extending from $x^1=-\infty$ to $x^1=+\infty$, with an 8-dimensional transverse space $\M_8$ within which the solution is spherically symmetric.

The solution has a charge as well, given by Gauss's law: $U=\int_{\partial\M_8}d^7\Sigma^mH_{m01}=6k\Omega_7$, where $\Omega_7$ is the volume of the unit 7-sphere corresponding to the infinite boundary of $\M_8$. This charge is equal to the ADM {\em tension} (energy/unit $x^1$ length) of the solution, so this string solution is an analogue of the extremal Reissner-Nordstrom solution of Einstein-Maxwell theory.

\begin{figure}[ht]
\begin{center}
\includegraphics[scale=.7]{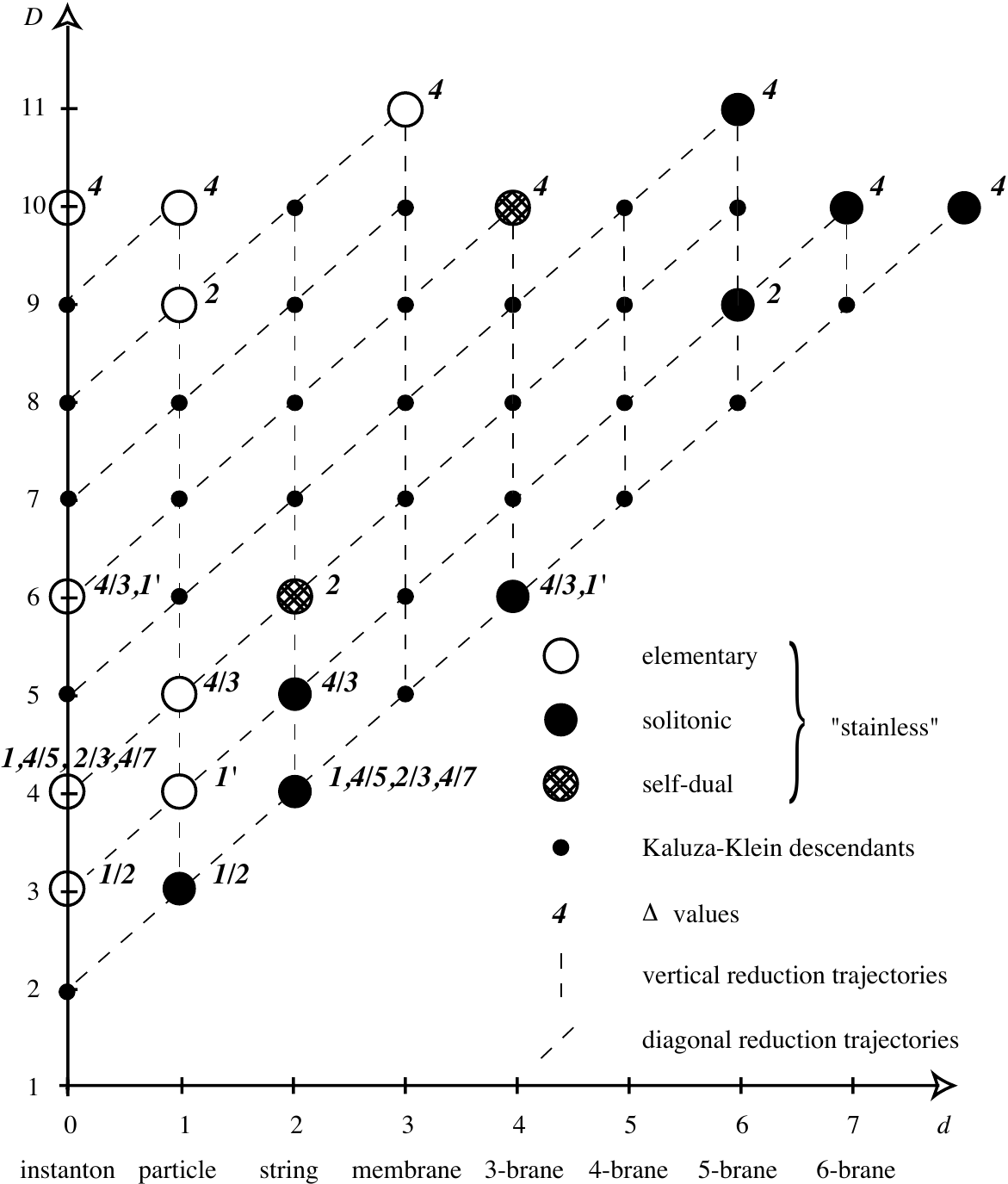}\\
\caption{``Brane-scan'' of supergravity $p$-brane solutions, linked by worldvolume (diagonal) and transverse-space (vertical) dimensional reductions}\label{branescan}
\end{center}
\end{figure}

There is a great variety of $p$-brane solutions in supergravity theories, of diverse worldvolume and transverse dimensionalities, as shown in Figure \ref{branescan}. The supersymmetric $p$-brane spectrum naturally generalizes the extremal black holes of Einstein-Maxwell theory, which may be viewed as 0-branes. In a given dimension of spacetime, the brane spectrum also naturally carries a representation of the corresponding supergravity {\em duality group}. 

Dimensional reduction of the maximal theory down from $D=11$ automatically generates a $GL(11-D,\bbR)$ nonlinearly realized symmetry of the $D$ dimensional reduced supergravity. However, special features of supergravity theories lead to an enhancement of this anticipated duality group. These features include the combination of vectors and scalars coming from the $D=11$ metric and the $D=11$ 3-form gauge field, and also the dualization of higher-rank form fields to lower-rank fields by Hodge dualization of the corresponding field strengths. 

The resulting duality groups for maximal supergravity are shown in Table \ref{dualitytab}, where $E_{11-D(11-D)}$ is the nonlinearly realized duality symmetry in spacetime dimension $D$, $K_D$ is its linearly realized maximal compact subgroup and $E_{11-D(11-D)}(\ZZ)$ is the discretized ``U-duality'' \cite{Hull:1994ys, Witten:1995ex} form consistent with the Dirac quantization condition, which is conjectured to survive in superstring theory.
\begin{table}[H]
  \centering
  \begin{tabular}{||c|c|c|c||}
\hline
$D$ & $E_{11-D(11-D)}(\mathbb R)$ & $K_D$ & $E_{11-D(11-D)}(\mathbb
Z)$\\
\hline
10A&$ \IR^+$& 1&1\\
10B&$ Sl(2,\IR)$&$SO(2)$& $Sl(2,\ZZ)$\\
9&$ Sl(2,\IR)\times \IR^+$& $SO(2)$&$Sl(2,\ZZ)$\\
8&$ Sl(3,\IR)\times Sl(2,\IR)$& $SO(3)\times SO(2)$&$Sl(3,\ZZ)\times Sl(2,\ZZ)$\\
7& $Sl(5,\IR)$ & $SO(5)$&$Sl(5,\ZZ)$\\
6& $ SO(5,5,\IR)$ & $SO(5)\times SO(5)$&$SO(5,5,\ZZ)$\\
5&$ E_{6(6)}(\IR)$& $USp(8)$&$E_{6(6)}(\ZZ)$\\
4&$ E_{7(7)}(\IR)$& $SU(8)/\ZZ_2$&$E_{7(7)}(\ZZ)$\\
3&$ E_{8(8)}(\IR)$& $SO(16)$&$E_{8(8)}(\ZZ)$\\
\hline
  \end{tabular}
  \caption{Supergravity $E_{11-D(11-D)}(\mathbb R)$ duality symmetries, $K_D$ maximal compact subgroups and the superstring $E_{11-D(11-D)}(\mathbb Z)$ discretizations\label{dualitytab}}
\end{table}

\section{The onset of supergravity divergences}
\label{sec:7}

Now we shall return to the initial question of field-theoretic gravity theories and their quantum problems. We have seen that there is a rich tapestry of supergravity limits, with surprizing additional duality symmetries, which emerge as ``zero-slope'' $\ap\to0$ limits of superstring/M-theory. The question remains whether links to theories based upon extended objects as the fundamental excitations have a bearing on the original ultraviolet problems of field-theoretic gravity and supergravity theories. The dimensional character of Newton's constant and the related nonlinearity of the Einstein-Hilbert action leads to a general expectation of nonrenormalizability in gravity and supergravity theories. Resolving the ultraviolet problem of quantum gravity has consequently been one of the main aims of superstring theory. However, it is important to understand precisely how the quantum properties of superstring theories differ from those of the corresponding supergravities when the latter are subject to standard field-theoretic quantization. 

In order to understand this relation, the precise order of onset of nonrenormalizable divergences in supergravity theories has remained an intensely studied question. Local supersymmetry brings about at least significant delays in the onset of ultraviolet divergences, but the full reach of the corresponding nonrenormalization theorems is still not fully clear. What is clear, however, is that links to superstring and M-theory have led to some genuinely surprizing ultraviolet cancellations.

Explicit calculations of ultraviolet divergence coefficients have been carried out using traditional Feynman diagram techniques up to the 2-loop level.\,\cite{Goroff:1985th, vandeVen:1991gw} Continuing on this way to higher loop orders, however, quickly becomes prohibitive: for the important 3-loop level at which the first dangerous counterterms occur, an estimate of the number of terms in a standard Feynman diagram calculation is of the order of $10^{20}$, owing to the complexity of the vertices and propagators.

Nonetheless, important progress using new techniques developed since 1998 has been made in the calculation of loop-diagram divergences in maximal supergravity and maximal super Yang-Mills theories.  These new methods use heavily the unitarity properties of Feynman diagrams, which generalize the optical theorem ${\rm Im} T =T^\ast T$  of ordinary quantum mechanics \cite{Bern:1998ug,Bern:2010ue,Bern:2012uf}.

Normally, one might think that one can only learn about the imaginary parts of quantum amplitudes using unitarity. However when the unitarity diagram cutting rules are combined with an expanded use of dimensional regularization, much more can be learned. In dimensional regularization, one analytically continues the dimension of spacetime in Feynman integrals away from the dimension of interest, \eg replacing $\int d^4k$ loop integrals by $\int d^{4+\epsilon}k$. 

The ordinary use of dimensional regularization focuses simply on the $\frac1\epsilon$ poles in quantum amplitudes, corresponding to logarithmic divergences in a straightforward high momentum cutoff regularization. However, one gets useful information by retaining the full $(4+\epsilon)$ dimensional amplitude. In such an analytically continued integral, an integrand $f(s)$ (where, \eg, $s$ is a Mandelstam momentum invariant, quadratic in loop momenta) will become deformed to $f(s)s^{-\epsilon/2}$ in order to balance dimensions. Then, since $s^{-\epsilon/2}=1-(\epsilon/2)\ln s + \ldots$ and since $\ln(s)=\ln(|s|)+i\pi\theta(s)$, one can learn about the real parts of an amplitude by retaining imaginary terms at order $\epsilon$.

The unitarity-based techniques allow for large classes of diagrams to become {\em cut constructible}, and allow for the eventual reduction of a higher-loop amplitude to integrals over products of tree amplitudes. At that point, other recent progress in the understanding of tree amplitudes comes into play. Although the individual Feynman diagrams at tree level are very complicated, sums of diagrams representing complete amplitudes can have striking simplicity, and in particular can satisfy powerful recursion relations \cite{Britto:2005fq}.

To date, these techniques have allowed the explicit calculation of maximal supergravity divergences to proceed up to the 4-loop level, something that would have been unthinkable using traditional Feynman diagram techniques \cite{Bern:2009kd}.
\begin{center}
\includegraphics[scale=.35]{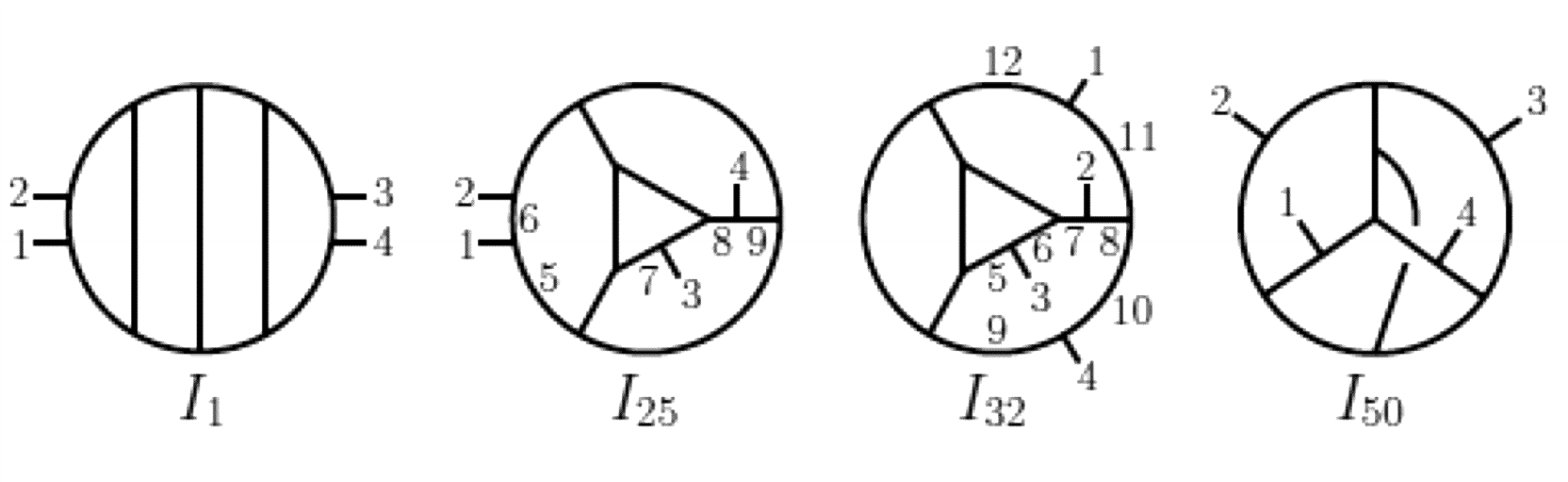}\\
\captionof{figure}{Four out of the 50 diagrams arising in the calculation of the 4-loop divergences in maximal supergravity}\label{4loopdiags}
\end{center}

The current status of the first possible maximal supergravity ultraviolet divergences is summarized in Table \ref{knowndivs}, with the currently known divergences shown in gray. The BPS degree represents the degree of supersymmetry invariance of the integrand prior to superspace integration. The surprizing feature of these results is the tardiness of maximal supergravity in getting around to revealing its ultraviolet divergences. At $D=8$ dimensions and $L=1$, one does indeed encounter the expected $R^4$ counterterm, but in lower spacetime dimensions this $\half$ BPS  divergence does not occur, leaving it to significantly less constrained counterterms to be the first UV divergence candidates.

\begin{table}[H]
 \centering
 \setlength{\tabcolsep}{3pt}
 \begin{tabular}{|l|c|c|c|c|c|c|c|}
\hline
Dimension $D$&11&10&8&7&6&5&4\\
\hline
Loop order $L$&\gray2&\gray2&\gray1&\gray2&\gray3&6?&5?\\
\hline
BPS degree\phantom{\Big|}&\gray0&\gray0&\gray$\ft12$&\gray$\ft14$&\gray$\ft18$&$0$&$\ft14$\\
\hline
Gen.\ form&\gray$\partial^{12} R^4$&\gray$\partial^{10} R^4$&\gray$R^4$&\gray$\partial^6 R^4$&\gray$\partial^6 R^4$&$\partial^{12} R^4$&$\partial^4 R^4$\\
\hline
 \end{tabular}
 \caption{Maximal supergravity first possible divergences \& BPS degree from unitarity-based calculations. Known divergences are shown in gray.\label{knowndivs}}
\end{table}

We will next see that careful analysis of the available counterterms reveals the reasons for the unanticipated divergence cancellations, and will push the anticipated first divergences even farther out than the calculational front shown in Table \ref{knowndivs}.

\subsection{Supergravity counterterm analysis}
\label{subsec:7.1}

The surprizing resilience of maximal $N=8$ supergravity to the threat of anticipated ultraviolet divergences has led to some speculation that perhaps superstring theory isn't actually necessary after all. Certainly, it has led to more than a decade of discussion between the unitarity-based calculators and supersymmetry practitioners in trying to understand what is going on. The current state of affairs reflects a significant deepening in understanding of the consequences of local supersymmetry and also of the r\^ole of duality symmetries of the maximal theory. 

In the 1980's, the understanding was that allowable counterterms would be those subject to nonrenormalization theorems based upon linearly realized supersymmetry, generalizing the famous nonrenormalization theorem that disallows as counterterms chiral superspace integrals (also called ``$F$ terms'') like $\int d^4x\, d^2\theta W(\phi)$ in $N=1$, $D=4$ supersymmetry, where $W(\phi)$ is a holomorphic function and $\phi$ is a chiral superfield satisfying $\bar D\phi=0$. Although such terms are fully allowed in a theory's classical action (and, indeed, play a critical r\^ole in supersymmetric extensions of the Standard Model), only full superspace integrals like $\int d^4x\, d^2\theta d^2\bar\theta\, K(\phi,\bar\phi)$ are allowed to occur as counterterms.

It was known by the mid 1980's that maximal supergravity and maximal super Yang-Mills theory could be quantized with at least half of their full supersymmetry manifestly linearly realized -- in a so-called ``off-shell'' formalism. This was explicitly constructed for the full maximal $N=4$ super Yang-Mills theory, but only for the linearized theory in the case of maximal $N=8$ supergravity. The resulting expectation was that the first allowed counterterms would have a $\int d^8\theta$ superspace integral structure in the case of maximal super Yang-Mills and a $\int d^{16}\theta$ structure in the case of maximal supergravity -- \ie full-superspace integrals for the linearly-realizable half supersymmetry.

Accordingly, the first allowed counterterms in maximal super Yang-Mills and maximal supergravity were considered to be \cite{Howe:1981xy, Kallosh:1980fi} of the following structures:
\begin{center}
\includegraphics[scale=.35]{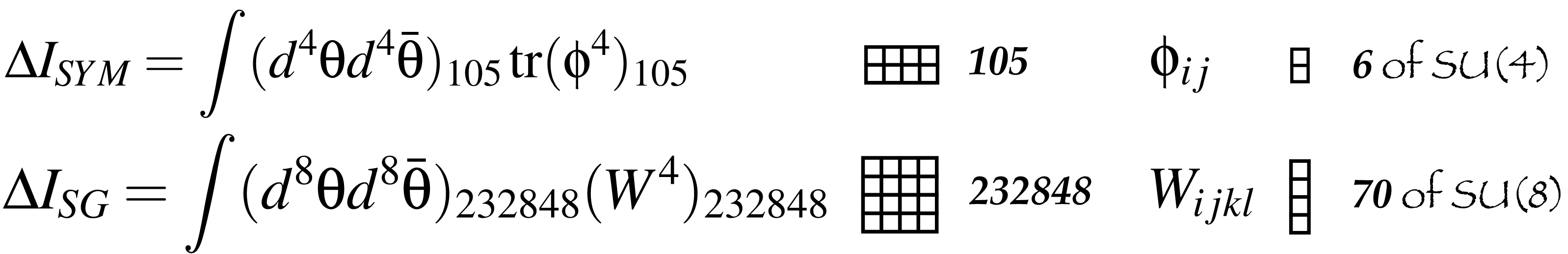}\\
\captionof{figure}{Half-BPS candidate counterterms for maximal super Yang-Mills and maximal supergravity theories}
\end{center}

Since $\int d^{16}\theta$ integration has dimension 16/2=8, one finds terms with  8 derivatives among the many terms produced by superspace integration in the maximal supergravity counterterm; general covariance requires these to be of the general form $\int R^4$. So the above understanding would just allow the dangerous 3-loop anticipated counterterm for the first $D=4$ divergence.

This expectation of $L=3$ loop first  divergences has clearly been upset by the unitarity-based calculations. This has led to a more detailed investigation of the nonrenormalization theorems, and in particular to use of the Ward identities for the full supersymmetry using a Batalin-Vilkovisky version of the BRST quantum formalism -- even though the transformations for the maximal theories are highly nonlinear and close to form an expected supersymmetry algebra only subject to the classical equations of motion. 

To do this requires adding source fields for a whole range of additional operators as needed to formulate the Ward identities. The resulting identities then take the form of a {\em cohomology problem} for a generalized exterior derivative, acting on triples of forms of adjacent rank. In addition, instead of computing beta functions for the coefficients of the expected counterterms, it is advantageous to consider operator insertions of the classical Lagrangian and its corresponding cocycle into its own quantum amplitudes, and then to calculate gamma functions for the allowed operator mixings of the Lagrangian cocycle with candidate counterterms and their cocycles, the latter being required to be consistent in structure with that of the Lagrangian cocycle \cite{Bossard:2009sy,Bossard:2009mn}.

In the case of maximal super Yang-Mills, such considerations are sufficient to explain all of the previously unanticipated single-trace operator cancellations. In the case of maximal supergravity, similar arguments show why the $R^4$ counterterm is in the end ruled out. But the unanticipated maximal supergravity cancellations found by the unitarity methods go much further than that: $D=5$, $L=2\,\&\,4$ and $D=4$, $L=3\,\&\,4$ divergences have all failed to occur.
 
In order to understand these further divergence cancellations, one needs to turn an apparent bug of the supergravity Ward identities into a feature. Their analysis is complicated by the density character of integrands in a locally supersymmetric theory. Counterterm cocycle component forms need to be pulled back to a bosonic, or ``body'' coordinate frame, and in local supersymmetry this involves not only the gravitational vielbeins, but also the fermionic gravitino fields.

To turn this density character to our advantage, one needs to combine it with the requirements of maximal supergravity's continuous duality invariance: $E_{7(7)}(\IR)$ in the $D=4$ case. For this purpose, one needs to know how to quantize while maintaining manifest duality symmetry, and this can be done in the case of the maximal $N=8$, $D=4$ theory by sacrificing manifest Lorentz invariance so as to handle the fact that only the $F_{\mu\nu}^i$ field strengths of the 28 vector fields can form an $E_{7(7)}(\IR)$ representation, separating them into self-dual and anti-self-dual parts in order to form a {\bf 56} of $E_{7(7)}(\IR)$. Sacrificing manifest Lorentz symmetry in this way, one can double the number of spacelike components of the vector fields, but reduce the number by a factor of $1/2$ by a duality constraint. In this formalism, one needs to check for the absence of anomalies in the now non-manifest Lorentz symmetry and also check the $SU(8)$ divisor subgroup of the duality group. Happily, these anomaly checks succeed, and as a result one may require that the perturbative field-theory counterterms preserve the continuous duality symmetry \cite{Bossard:2010dq}.

Combining the requirements of the full local supersymmetry Ward identities with the requirement of continuous duality symmetry, all of the outstanding divergence cancellations currently found by the unitarity methods have been explained, and new cancellations are now predicted at $D=4$, $L=5\,\&\,6$. The resulting pattern of anticipated\footnote{
The $D=4$, $L=7$ situation requires special care \cite{Bossard:2011tq}. At the linearized level, it would seem that the $\partial^8 R^4$ candidate could be the first full-superspace non-BPS counterterm in $D=4$. The volume of superspace, $\int d^4x d^{32}\theta \det(E_M^A)$ would seem to be the obvious candidate. However, rather surprisingly, it turns out that this superspace volume vanishes subject to the classical field equations, so there is no need for such a counterterm in the renormalized action. Instead, what looks like a non-BPS $\partial^8 R^4$ counterterm at the linearized level turns into a $\frac18$-BPS counterterm at the full nonlinear level. This illustrates the important r\^ole that nonlinear structure can play in quantum gravity divergence analysis.
} first divergences is shown in Table \ref{anticipdivs}.
\begin{table}[H]
 \centering\setlength{\tabcolsep}{3pt}
 \begin{tabular}{|l|c|c|c|c|c|c|c|}
\hline
Dimension $D$&11&10&8&7&6&5&4\\
\hline
Loop order $L$&{\gray2}&{\gray2}&{\gray1}&{\gray2}&{\gray3}&6&7\\
\hline
BPS degree\phantom{\Big|}&{\gray0}&{\gray0}&{\gray$\ft12$}&{\gray$\ft14$}&{\gray$\ft18$}&$0$&$\ft18$\\
\hline
Gen.\ form&{\gray$\partial^{12} R^4$}&{\gray$\partial^{10} R^4$}&{\gray$R^4$}&{\gray$\partial^6 R^4$}&{\gray$\partial^6 R^4$}&$\partial^{12} R^4$&$\partial^8 R^4$\\
\hline
 \end{tabular}
\caption{Maximal supergravity current first divergence expectations \& BPS degree. Known divergences are shown in gray and the first anticipated $D=4$ and $D=5$ divergences are shown in black.\label{anticipdivs}}
\end{table}

\subsection{Supergravity divergences from superstrings}
\label{subsec:7.2}

A satisfying aspect of the current understanding of the relation between superstring theory and maximal supergravity is that one now obtains exactly the above predictions for the onset of supergravity divergences from a superstring perspective as well. In this regard, one may view superstring theory as a rather elaborate ``regulator'' of the supergravity quantum amplitudes. 

In the critical dimension $D=10$ for superstrings, the $R^4$ correction to the effective field-theory Lagrangian occurs with a coefficient ${\ap}^3$, as can be seen on dimensional grounds, since $\ap$ has dimensions of $(\hbox{length})^2$, so ${\ap}^3R^4$ has the same dimensions as $R$. How then can there be divergences in the effective field theory, which is obtained by taking the limit $\ap\rightarrow0$?

Recall that in order to compare $D=4$ maximal supergravity to the string-theory effective action, one must dimensionally reduce on a torus $T^6$. Start from the string frame, in which the gravitational Lagrangian $e^{-2\phi}R$ has a scalar prefactor whose v.e.v. $g_s^{-2}$ gives a $D=10$ Newton's constant $G_{10}\sim g_s^2\ell_s^8$, where $\ell_s$ is the string scale, needed on dimensional grounds and related to $\ap$ by $\ap\sim\frac{\ell_s^2}{2\hbar^2c^2}$; $g_s$ is the string coupling constant. If the typical scale of one of the compact toroidal dimensions is $\R$, then reduction of the effective action on $T^6$ produces an extra prefactor of $\R^6$ in the $D=4$ effective action, giving a $D=4$ Newton's constant $G_4\sim \frac{g_s^2\ell_s^8}{\R^6}$. Thus, the $D=4$ Plank length $\ell_4\sim (G_4)^{\half}$ is related to the string coupling constant by $g_s\sim \frac{\R^3}{\ell_s^4}\ell_4$.

In order to compare the dimensionally reduced string effective action to quantized $D=4$  maximal supergravity, one needs to ensure that the $D=4$ Newton's constant remains finite, while the towers of excited string states and also the Kaluza-Klein excitations from the dimensional reduction all have masses that are infinitely large compared to the $D=4$ Planck scale $\ell_4$. This is achieved by taking $\frac1{\R}\,,\,\frac1{\ell_s}\,\&\,\frac{\R}{\ell_s^2}\gg\frac1{\ell_4}$\,, which is compatible with holding $g_s$ and $\ell_4$ fixed while taking $\ell_s\sim\R^{\frac34}(\frac{\ell_4}{g_s})^{\frac14}\rightarrow0$. 

This analysis would seem to indicate that the effective field theory could be ultraviolet finite. However, this string analysis can be misleading because string nonperturbative effects can conspire to swamp what one might otherwise want to identify as the field-theoretic supergravity contribution \cite{Green:2007zzb}. Analysis of this ``decoupling'' problem in the above supergravity limit shows that decoupling can be carried out for loop orders $L\le 6$, but that beyond that order the decoupling issues prevent conclusions about field-theoretic finiteness from being made \cite{Green:2010sp}.
Moreover, analysis of the superstring effective action contributions also shows they can be continuously $E_{7(7)}(\IR)$ invariant to the same order \cite{Elvang:2010kc, Beisert:2010jx}.

\section{Other aspects of string theory}
\label{sec:8}

\subsection{The String Scale}
\label{subsec:8.1}

Let us now consider the physically relevant energy or inverse-length scales characterizing the higher dimensions inherent in string theory. The analysis will be similar to that of Section \ref{subsec:7.2}, except that instead of looking at the field-theory limit $\ell_s\to 0$, we now consider the {\em finite} values of $\ell_s$ that are compatible with perturbatively realistic particle physics. First consider the context of a traditional Kaluza-Klein reduction, starting in $D=10$ with $I_{\sst{10}}=\ell_s^{-8}\int d^{10}x\,e^{-2\phi}(R+\ell_s^2F^2)$, where $F$ represents the Yang-Mills field strength. Next, dimensionally reduce down to $D=4$ on a manifold of volume $V_6$ while replacing $\phi$ by $\phi_0=\langle\phi\rangle$. The $D=4$ reduced action then becomes
\be
I_{\sst4}=\frac{V_6}{\ell_s^8}e^{-2\phi_0}\int d^4x(R+\ell_s^2F^2)\ .\nn
\ee
so in $D=4$ we can identify 
\be
M_{\rm\sst{Pl}}=\frac{V_6^{\half}}{{\ell_s}^4}e^{-\phi_0}\qquad\hbox{and}\qquad g_{\rm\sst{YM}}=\frac{e^{\phi_0}\ell_s^3}{V_6^{\half}}\ .\nn
\ee
Now, to avoid strong coupling in the $D=10$ string theory, one requires $e^{\phi_0}<1$ while in $D=4$, $g^2_{\rm\sst{YM}}\sim\frac1{30}$. Hence $V_6\ell_s^{-6}=e^{2\phi_0}g_{\rm\sst{YM}}^{-2}\le30$ and so for $V_6\sim \R^6$, one finds $\ell_s\sim \R$. Moreover, substituting for $e^{-\phi_0}$ in terms of $g_{\rm\sst{YM}}$, one finds $M_{\rm\sst{Pl}}=(\ell_s g_{\rm\sst{YM}})^{-1}$, requiring $\ell_s\sim\frac{10}{M_{\rm{Pl}}}$. This is as one might expect: in the standard dimensional reduction scenario, intrinsic string-theory effects cannot occur too far below the Planck scale.

Consider now how the analysis changes when one switches from the standard Kaluza-Klein reduction to a braneworld scenario. In a braneworld scenario, one proposes that the observable lower-dimensional universe is concentrated on a subsurface of the higher-dimensional spacetime, instead of being smeared evenly over the extra dimensions, \ie instead of assuming there is no dependence on the extra coordinates.

To see how the string-scale analysis changes, consider a ${\rm D}_p$-brane with $p>3$ and with gauge fields defined only on the $d=p+1$ dimensional D-brane worldvolume. Of the $p$ spatial D-brane dimensions,  $p-3$ are compactified and the remaining $3$ coincide with the spatial directions of the reduced $D=4$ spacetime. The starting action is now $\ell_s^{-8}\int d^{10}x\, e^{-2\phi} R + \ell_s^{3-p}\int d^{p+1}\xi\, e^{-\phi}F^2$\,. As before, one has $M_{\rm\sst{Pl}}=V_6^{\half}\ell_s^{-4}e^{-\phi_0}$ but now $g_{\rm\sst{YM}}=e^{\half\phi_0}\ell_s^{\half(p-3)}{V^{-\half}_{p-3}}$.

Limiting again the string coupling to perturbative values $e^{\phi_0}\le1$, find $V_{p-3}\ell_s^{3-p}\le g^{-2}_{\rm\sst{YM}}\sim 30$. Now, however, for $V_6=V_{p-3}V_{9-p}$ one can have {\em independent} $\ell_{\rm p}=V^{\frac1{p-3}}_{p-3}$ and $\R=V^{\frac1{9-p}}_{9-p}$.
One may  write $M^2_{\rm\sst{Pl}}=\ell_s^{-2}(V_{p-3}\ell_s^{3-p})(V_{9-p}\ell_s^{p-9})e^{-2\phi_0}$. Then from $V_{p-3}\ell_s^{3-p}\le 30$, one learns $\ell_s\sim\ell_{\rm p}(30)^{-\frac1{p-3}}$, while if $\R\gg\ell_{\rm p}$ (giving a highly asymmetric $V_6$), one has $V_{9-p}\ell_s^{p-9}\gg1$. Thus $\ell_s^2\gg M_{\rm\sst{pl}}^{-2}$ is now possible, \ie one can have a string mass scale {\em significantly below} the Planck scale \cite{Antoniadis:1998ig}.

Such brane-world considerations lie at the base of current experimental protocols searching for string or gravitational phenomena at the Large Hadron Collider at CERN. A low string mass scale corresponds also to a low scale for inherently quantum-gravitational effects such as the threshold production of black holes. These would immediately then decay by Hawking radiation, but the resulting resonance could have a striking new experimental signature.

\subsection{Boundaries of moduli space}
\label{subsec:8.2}

The infinities of perturbative field theory are tamed by string theory. But the story of infinities does not end there. In string theory, the most important singularities occur at the {\em boundaries of moduli space}, \eg in amplitudes where moduli are about to pinch off so that topology change can take place in a Riemann surface.

\begin{figure}[ht]
\begin{center}
\includegraphics[scale=.7]{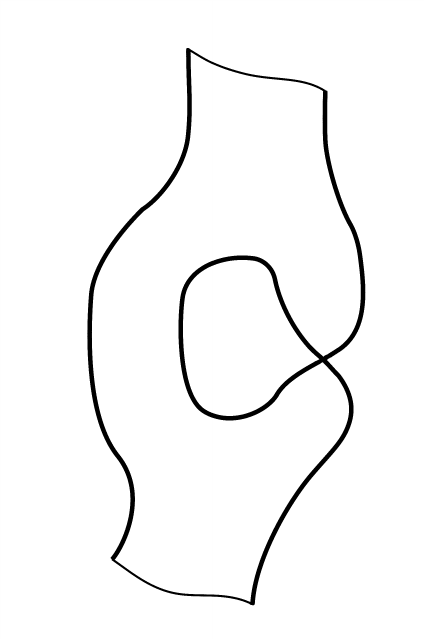}\\
\caption{Topology change at a boundary of moduli space}
\end{center}
\end{figure}

These boundary configurations are the places where one really has to check whether string theory is free of infinities. And often, when divergences seem apparent, one realizes that the singularities should be ``blown up'', with new modes particular to the singularity structure of the Riemann surface appearing. These can contribute to blown-up singularity sectors with new gauge fields and other massless modes in the effective field theory, so such special points in moduli space can also give rise to important physical effects at the same time as posing a challenge to the analysis of divergences.

The calculation of higher-loop string amplitudes requires detailed control of the Riemann surface moduli in the quantum path integral. This has been comprehensively done up through the two-loop order \cite{D'Hoker:2001nj,D'Hoker:2001it,D'Hoker:2001it,D'Hoker:2001qp}. This has enabled proofs of the finiteness of string theory up to this level to be given \cite {Mandelstam:1991tw,Berkovits:1993hg}. More recent developments have employed a pure-spinor approach to the string worldsheet \cite{Berkovits:1994wr,Berkovits:1999in} are expected to lead to an all-orders resolution of this key question.

\subsection{String and Gravity Thermodynamics}
\label{subsec:8.3}

One of the most famous results of string theory has been the derivation of the Bekenstein-Hawking formula $S=\frac{A}{4G}$ for the entropy of a black hole in terms of the area $A$ of the horizon. This derivation employed nearly supersymmetric (\ie nearly BPS) configurations in order to enable a detailed microstate counting agreeing with the Bekenstein-Hawking entropy formula \cite{Strominger:1996sh}. Current work includes study of the deviation from a blackbody to a ``greybody'' spectrum in the emitted Hawking radiation.

Related work is aimed at understanding whether string theory evades Hawking's prediction that black holes lead to a loss of quantum information. This has led to ``fuzzball'' formulations of macroscopic blackholes in terms of an average over BPS coherent states describing individual nonsingular geometries, in order to give an account of the thermodynamics of non-supersymmetric black holes.

\begin{figure}[ht]
\begin{center}
\includegraphics[scale=.4]{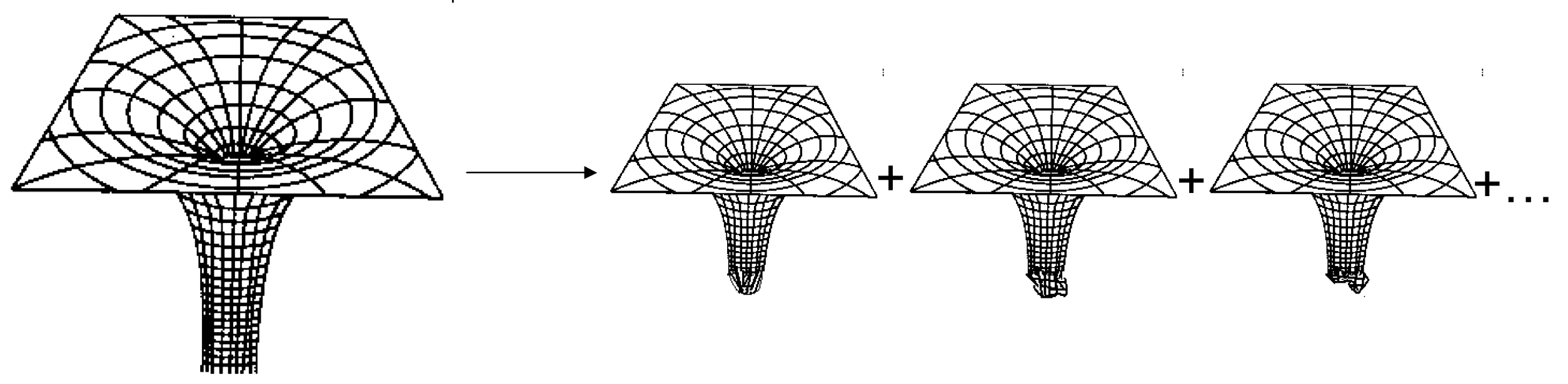}\\
\caption{``Fuzzball'' effective geometry from an average over nonsingular geometries}
\end{center}
\end{figure}

\section{Conclusion}
\label{sec:9}

With the expansion of string theory to the M-theory picture of a web of perturbatively-defined theories (plus $D=11$ supergravity) linked by duality symmetries, a breathtaking unification vista has opened. And yet, this unification remains largely an aspiration: despite many striking confirmations of duality relations, a proper definition of the fundamental states of M-theory and a corresponding derivation of the duality symmetries acting on them remain to be given. Considering the history of nonperturbative relations in field theory, this should not be so surprizing. Many of the currently most important theoretical constructions (Yang-Mills theory, supersymmetry, even $\phi^4$ theory) are not rigorously formulated outside perturbation theory. Perhaps M-theory will upend this situation and in the end become better formally grounded than phenomenological field theory currently is.

The theme of duality, meanwhile, continues to make further surprizing connections between previously distinct subjects. The picture of stacks of $(3+1)$-dimensional D-branes in type IIB theory, with braneworld Goldstone modes constituting vector supermultiplets and interactions via open superstrings linking the D-brane stacks to form nonabelian gauge theories, forms a ``holomorphic'' construction of supersymmetric Yang-Mills theory on the boundary of a $(4+1)$-dimensional spacetime. This ``gauge-gravity duality'' provides powerful nonperturbative relations on both the ``gravity'' (or superstring) and ``gauge'' sides of the correspondence (for a review, see \cite{Aharony:1999ti}.)

The gauge-gravity correspondence has evolved to cover many situations where nonperturbative gauge-theory phenomena on boundaries are important. Key applications currently include the dynamics of quark-gluon plasmas, Navier-Stokes fluid hydrodynamics, and applications in high-temperature superconductivity. The gauge-gravity duality constructions can provide information about phases where the usual field-theoretic partition function is neither convergent not Borel-summable: in other words, in deeply nonperturbative regions of phase space.

Historically, string theory was born out of the study of resonances in strong-interaction physics. With some of the recent developments, it may in a sense be returning to its birth context. But it also remains the strongest current approach towards solving the key remaining problem of theoretical physics, which is the formulation of a quantum theory of gravity.


\begin{thebibliography}{99.}%

\bibitem{Goroff:1985th}
  M.~H.~Goroff and A.~Sagnotti,
  ``The Ultraviolet Behavior of Einstein Gravity,''
  Nucl.\ Phys.\ B {\bf 266} (1986) 709.

\bibitem{vandeVen:1991gw}
  A.~E.~M.~van de Ven,
  ``Two loop quantum gravity,''
  Nucl.\ Phys.\ B {\bf 378} (1992) 309.

\bibitem{Deser:1977nt}
  S.~Deser, J.~H.~Kay and K.~S.~Stelle,
  ``Renormalizability Properties of Supergravity,''
  Phys.\ Rev.\ Lett.\  {\bf 38} (1977) 527.
  
\bibitem{Deser:1976rb}
  S.~Deser and B.~Zumino,
  ``A Complete Action for the Spinning String,''
  Phys.\ Lett.\ B {\bf 65} (1976) 369.
  
\bibitem{Brink:1976sc}
  L.~Brink, P.~Di Vecchia and P.~S.~Howe,
  ``A Locally Supersymmetric and Reparametrization Invariant Action for the Spinning String,''
  Phys.\ Lett.\ B {\bf 65} (1976) 471.
  
\bibitem{Polyakov:1981rd}
  A.~M.~Polyakov,
  ``Quantum geometry of bosonic strings,''
  Phys.\ Lett.\  B {\bf 103} (1981) 207.

\bibitem{Callan:1985ia}
  C.~G.~Callan, Jr., E.~J.~Martinec, M.~J.~Perry and D.~Friedan,
  ``Strings in Background Fields,''
  Nucl.\ Phys.\ B {\bf 262} (1985) 593.
  
\bibitem{Buscher:1987sk}
  T.~H.~Buscher,
  ``A Symmetry of the String Background Field Equations,''
  Phys.\ Lett.\ B {\bf 194} (1987) 59.
  
\bibitem{Dai:1989ua}
  J.~Dai, R.~G.~Leigh and J.~Polchinski,
  ``New Connections Between String Theories,''
  Mod.\ Phys.\ Lett.\ A {\bf 4} (1989) 2073.
  
\bibitem{Dine:1989vu}
  M.~Dine, P.~Y.~Huet and N.~Seiberg,
  ``Large and Small Radius in String Theory,''
  Nucl.\ Phys.\ B {\bf 322} (1989) 301.
  
\bibitem{Cvetic:1999zs}
  M.~Cvetic, H.~Lu, C.~N.~Pope and K.~S.~Stelle,
  ``T duality in the Green-Schwarz formalism, and the massless / massive IIA duality map,''
  Nucl.\ Phys.\ B {\bf 573} (2000) 149
  [hep-th/9907202].
  
\bibitem{Hull:1994ys}
  C.~M.~Hull and P.~K.~Townsend,
  ``Unity of superstring dualities,''
  Nucl.\ Phys.\ B {\bf 438} (1995) 109
  [hep-th/9410167].
  
\bibitem{Witten:1995ex}
  E.~Witten,
  ``String theory dynamics in various dimensions,''
  Nucl.\ Phys.\ B {\bf 443} (1995) 85
  [hep-th/9503124].
  
\bibitem{Fradkin:1985ys}
  E.~S.~Fradkin and A.~A.~Tseytlin,
  ``Quantum String Theory Effective Action,''
  Nucl.\ Phys.\ B {\bf 261} (1985) 1.
  
\bibitem{Bern:1998ug}
  Z.~Bern, L.~J.~Dixon, D.~C.~Dunbar, M.~Perelstein and J.~S.~Rozowsky,
  ``On the relationship between Yang-Mills theory and gravity and its implication for ultraviolet divergences,''
  Nucl.\ Phys.\ B {\bf 530} (1998) 401
  [hep-th/9802162].

\bibitem{Bern:2010ue}
  Z.~Bern, J.~J.~M.~Carrasco and H.~Johansson,
  ``Perturbative Quantum Gravity as a Double Copy of Gauge Theory,''
  Phys.\ Rev.\ Lett.\  {\bf 105} (2010) 061602
  [arXiv:1004.0476 [hep-th]].

\bibitem{Bern:2012uf}
  Z.~Bern, J.~J.~M.~Carrasco, L.~J.~Dixon, H.~Johansson and R.~Roiban,
  ``Simplifying Multiloop Integrands and Ultraviolet Divergences of Gauge Theory and Gravity Amplitudes,''
  arXiv:1201.5366 [hep-th].
  
\bibitem{Britto:2005fq}
  R.~Britto, F.~Cachazo, B.~Feng and E.~Witten,
  ``Direct proof of tree-level recursion relation in Yang-Mills theory,''
  Phys.\ Rev.\ Lett.\  {\bf 94} (2005) 181602
  [hep-th/0501052].
  
\bibitem{Bern:2009kd}
  Z.~Bern, J.~J.~Carrasco, L.~J.~Dixon, H.~Johansson and R.~Roiban,
  ``The Ultraviolet Behavior of N=8 Supergravity at Four Loops,''
  Phys.\ Rev.\ Lett.\  {\bf 103} (2009) 081301
  [arXiv:0905.2326 [hep-th]].
  
\bibitem{Howe:1981xy}
  P.~S.~Howe, K.~S.~Stelle and P.~K.~Townsend,
  ``Superactions,''
  Nucl.\ Phys.\ B {\bf 191} (1981) 445.
  
\bibitem{Kallosh:1980fi}
  R.~E.~Kallosh,
  ``Counterterms in extended supergravities,''
  Phys.\ Lett.\ B {\bf 99} (1981) 122.
  
\bibitem{Bossard:2009sy}
  G.~Bossard, P.~S.~Howe and K.~S.~Stelle,
  ``The Ultra-violet question in maximally supersymmetric field theories,''
  Gen.\ Rel.\ Grav.\  {\bf 41} (2009) 919
  [arXiv:0901.4661 [hep-th]].
  
\bibitem{Bossard:2009mn}
  G.~Bossard, P.~S.~Howe and K.~S.~Stelle,
  ``A Note on the UV behaviour of maximally supersymmetric Yang-Mills theories,''
  Phys.\ Lett.\ B {\bf 682} (2009) 137
  [arXiv:0908.3883 [hep-th]].

\bibitem{Bossard:2010dq}
  G.~Bossard, C.~Hillmann and H.~Nicolai,
  ``E7(7) symmetry in perturbatively quantised N=8 supergravity,''
  JHEP {\bf 1012} (2010) 052
  [arXiv:1007.5472 [hep-th]].
\bibitem{Bossard:2011tq}
  G.~Bossard, P.~S.~Howe, K.~S.~Stelle and P.~Vanhove,
  ``The vanishing volume of D=4 superspace,''
ÊÊClass.\ Quant.\ Grav.\  {\bf 28} (2011) 215005
ÊÊ[arXiv:1105.6087 [hep-th]].
ÊÊ

\bibitem{Green:2007zzb}
  M.~B.~Green, H.~Ooguri and J.~H.~Schwarz,
  ``Nondecoupling of Maximal Supergravity from the Superstring,''
ÊÊPhys.\ Rev.\ Lett.\  {\bf 99} (2007) 041601
ÊÊ[arXiv:0704.0777 [hep-th]].
ÊÊ
  
\bibitem{Green:2010sp}
  M.~B.~Green, J.~G.~Russo and P.~Vanhove,
  ``String theory dualities and supergravity divergences,''
  JHEP {\bf 1006} (2010) 075
  [arXiv:1002.3805 [hep-th]].
  
\bibitem{Elvang:2010kc}
  H.~Elvang and M.~Kiermaier,
  ``Stringy KLT relations, global symmetries, and $E_{7(7)}$ violation,''
  JHEP {\bf 1010} (2010) 108
  [arXiv:1007.4813 [hep-th]].
  
\bibitem{Beisert:2010jx}
  N.~Beisert, H.~Elvang, D.~Z.~Freedman, M.~Kiermaier, A.~Morales and S.~Stieberger,
  ``E7(7) constraints on counterterms in N=8 supergravity,''
  Phys.\ Lett.\ B {\bf 694} (2010) 265
  [arXiv:1009.1643 [hep-th]].
  
\bibitem{Antoniadis:1998ig}
  I.~Antoniadis, N.~Arkani-Hamed, S.~Dimopoulos and G.~R.~Dvali,
  ``New dimensions at a millimeter to a Fermi and superstrings at a TeV,''
  Phys.\ Lett.\ B {\bf 436} (1998) 257
  [hep-ph/9804398].
  

\bibitem{D'Hoker:2001nj}
  E.~D'Hoker and D.~H.~Phong,
  ``Two loop superstrings. 1. Main formulas,''
  Phys.\ Lett.\ B {\bf 529} (2002) 241
  [hep-th/0110247].
  
\bibitem{D'Hoker:2001zp}
  E.~D'Hoker and D.~H.~Phong,
  ``Two loop superstrings. 2. The Chiral measure on moduli space,''
  Nucl.\ Phys.\ B {\bf 636} (2002) 3
  [hep-th/0110283].

\bibitem{D'Hoker:2001it}
  E.~D'Hoker and D.~H.~Phong,
  ``Two loop superstrings. 3. Slice independence and absence of ambiguities,''
  Nucl.\ Phys.\ B {\bf 636} (2002) 61
  [hep-th/0111016].
  
\bibitem{D'Hoker:2001qp}
  E.~D'Hoker and D.~H.~Phong,
  ``Two loop superstrings 4: The Cosmological constant and modular forms,''
  Nucl.\ Phys.\ B {\bf 639} (2002) 129
  [hep-th/0111040].
  
\bibitem{Mandelstam:1991tw}
  S.~Mandelstam,
  ``The n loop string amplitude: Explicit formulas, finiteness and absence of
  ambiguities,''
  Phys.\ Lett.\  B {\bf 277} (1992) 82.

\bibitem{Berkovits:1993hg}
  N.~Berkovits,
  ``Finiteness and unitarity of Lorentz covariant Green-Schwarz superstring amplitudes,''
  Nucl.\ Phys.\ B {\bf 408} (1993) 43
  [hep-th/9303122].
  
\bibitem{Berkovits:1994wr}
  N.~Berkovits,
  ``Covariant quantization of the Green-Schwarz superstring in a Calabi-Yau background,''
  Nucl.\ Phys.\ B {\bf 431} (1994) 258
  [hep-th/9404162].
  
\bibitem{Berkovits:1999in}
  N.~Berkovits,
  ``Quantization of the superstring with manifest U(5) superPoincare invariance,''
  Phys.\ Lett.\ B {\bf 457} (1999) 94
  [hep-th/9902099].
  
\bibitem{Strominger:1996sh}
  A.~Strominger and C.~Vafa,
  ``Microscopic origin of the Bekenstein-Hawking entropy,''
  Phys.\ Lett.\ B {\bf 379} (1996) 99
  [hep-th/9601029].
\bibitem{Aharony:1999ti}
  O.~Aharony, S.~S.~Gubser, J.~M.~Maldacena, H.~Ooguri and Y.~Oz,
  ``Large N field theories, string theory and gravity,''
ÊÊPhys.\ Rept.\  {\bf 323} (2000) 183
ÊÊ[hep-th/9905111].
ÊÊ
  
\end{thebibliography}
\end{document}